\documentclass[11pt]{article}
\usepackage{a4wide}
\usepackage{amssymb}
\usepackage{graphicx}

\usepackage{amsmath}
\usepackage{amsfonts}

\usepackage{amsthm}

\usepackage{stmaryrd}

\usepackage{enumerate}

\usepackage{url}

\usepackage{pj-macros}

\begin{document}

\title{Deciding semantic finiteness of
	pushdown processes and
	first-order grammars {w.r.t.}~bisimulation equivalence
}

\author{Petr Jan\v{c}ar\\
	{\small	Dept of Computer Science, Faculty of Science, Palack\'{y}
	Univ., Olomouc, Czechia}
}

\date{}

\maketitle

\begin{abstract}
The problem if a given configuration  of a pushdown automaton (PDA) is
bisimilar with some (unspecified) finite-state process is shown to be
decidable. The decidability is proven in the framework of first-order grammars, which are given by finite sets 
of labelled rules that rewrite roots of first-order terms.
	The framework is equivalent to PDA where also deterministic
	({i.e.}~alternative-free) 
 epsilon-steps are allowed, hence to the model for which S\'enizergues
 showed an involved procedure deciding bisimilarity (1998, 2005). 
Such a procedure is here used as a black-box part of the algorithm.

The result extends the decidability of the regularity problem
for deterministic PDA that was shown by Stearns (1967), and later
improved  by Valiant (1975) regarding the complexity.	
The decidability question for nondeterministic PDA,
answered positively here, had been open 
(as indicated, e.g., by Broadbent and G\"oller, 2012). 
\end{abstract}

\section{Introduction}\label{sec:intro}

The question of deciding 
semantic equivalences of systems, like language equivalence,
has been a frequent topic in
computer science. A closely related question asks if 
a given system
in a class $\calC_1$
has an equivalent in a subclass $\calC_2$.
Pushdown automata (PDA) 
constitute a well-known example;
language equivalence 
and regularity 
are undecidable for PDA. 
In the case of deterministic PDA (DPDA), the decidability and
complexity results for 
regularity~\cite{DBLP:journals/iandc/Stearns67,DBLP:journals/jacm/Valiant75}
preceded the famous
decidability result for equivalence 
by S\'enizergues~\cite{Senizergues:TCS2001}.

In concurrency theory, logic, verification, and other areas,
a finer equivalence, called \emph{bisimulation equivalence} or
\emph{bisimilarity}, has emerged as another fundamental behavioural
equivalence (cf., e.g.,~\cite{Milner1989}); on deterministic systems it essentially coincides with
language equivalence.
An on-line survey of the results which study this equivalence 
in a specific area of process
rewrite systems is maintained by Srba~\cite{Srba:Roadmap:04}.

One of the most involved results in this area is the decidability of bisimilarity
for pushdown processes
generated by (nondeterministic) PDA in which 
$\varepsilon$-steps are restricted so that each $\varepsilon$-step
has no alternative (and can be restricted
to be popping); 
this result was shown by S\'enizergues~\cite{Seni05} who
thus generalized his above mentioned result for DPDA.
There is no known upper bound on the complexity 
of this decidable problem. 
The nonelementary lower bound established in~\cite{BGKM12}
is, in fact, TOWER-hardness in the terminology
of~\cite{DBLP:journals/toct/Schmitz16}, and it holds
even for real-time PDA, {i.e.}~PDA with no $\varepsilon$-steps. 
For the above mentioned PDA with restricted 
$\varepsilon$-steps the bisimilarity problem is even not primitive
recursive;
its Ackermann-hardness is shown in~\cite{DBLP:conf/fossacs/Jancar14}.
In the deterministic case, the equivalence problem is known to be
PTIME-hard, and
has a primitive recursive upper bound shown 
by Stirling~\cite{Stir:DPDA:prim} (where a finer analysis places
the problem in TOWER~\cite{DBLP:conf/fossacs/Jancar14}).

Extrapolating the deterministic case,
we might expect that for PDA the ``regularity'' problem
{w.r.t.}~bisimilarity (asking if a given PDA-configuration is bisimilar
with a state in a finite-state system)
is decidable as well,
and that this problem might be easier
than the equivalence problem solved in~\cite{Seni05}; only
EXPTIME-hardness is known here (see~\cite{Kucera10}, and
\cite{Srba:Roadmap:04} for detailed
references).
Nevertheless, this decidability question has been open so far, as also
indicated in~\cite{DBLP:conf/fsttcs/BroadbentG12} 
(besides \cite{Srba:Roadmap:04}).

\emph{Contribution of this paper.}
We show that semantic finiteness of pushdown configurations
{w.r.t.}~bisimilarity is decidable.
The decidability is proven in the framework of \emph{first-order
grammars},
{i.e.}~of finite sets 
of labelled rules that rewrite roots of first-order terms.
Though we do not use (explicit) $\varepsilon$-steps, the framework 
is equivalent to the model of PDA with restricted $\varepsilon$-steps
for which 
S\'enizergues's general decidability proof~\cite{Seni05} applies.
(A simplified proof directly in the first-order grammar
framework, hence an alternative to the proof in~\cite{Seni05},
is given in~\cite{DBLP:conf/icalp/Jancar14}.)

The presented
algorithm, answering if a given configuration, {i.e.}~a first-order term
$E_0$
in the labelled transition system generated by a first-order grammar, 
has a bisimilar finite-state
system, uses the result of~\cite{Seni05} 
(or of~\cite{DBLP:conf/icalp/Jancar14}) as a
black-box procedure. 
By~\cite{DBLP:conf/fossacs/Jancar14} we cannot get  
a primitive recursive upper bound via a black-box use of the
decision procedure for bisimilarity.

Semidecidability of the semantic finiteness
problem has been long clear, hence it is
the existence of finite effectively verifiable witnesses of 
the negative case
that is the crucial point here. 
It turns out that a witness of semantic infiniteness 
of a term (i.e., of a configuration) $E_0$ is a specific path  $E_0\gt{u}\gt{w}$
in the respective labelled transition system where the sequence $w$ of
actions can be repeated
forever. 
The idea how to verify if the respective infinite path 
$E_0\gt{u}\gt{w}\gt{w}\gt{w}\cdots$, denoted 
$E_0\gt{u}\gt{w^\omega}$,
visits
terms (configurations) from infinitely many equivalence classes is to
consider the ``limit term'' $\LIMIT$ that is ``reached'' by
$E_0\gt{u}\gt{w^\omega}$; the term $\LIMIT$ is generally infinite but 
regular 
(i.e., it has only finitely many subterms).
The (black-box) procedure deciding equivalence is used for computing
a finite number $e$ such that we are guaranteed that if  
$E_0\gt{u}\gt{w^e}$ does not reach a term equivalent to $\LIMIT$ then 
$E_0\gt{u}\gt{w^k}$ does not reach such a term for any $k\geq e$. 
In this case the path $E_0\gt{u}\gt{w^\omega}$ indeed visits terms in
infinitely many equivalence classes since the visited terms approach $\LIMIT$
syntactically and thus also semantically (by increasing the 
``equivalence-level'' with $\LIMIT$) but never belong to the equivalence
class of $\LIMIT$.
To show the existence of a respective witness $E_0\gt{u}\gt{w}$ for each 
semantically infinite $E_0$ is not trivial but it can done by a
detailed study
of the paths $E_0\gt{a_1}E_1\gt{a_2}E_2\gt{a_3}\cdots$ where $E_i$ are
from pairwise different equivalence classes; here we also use the
infinite Ramsey theorem for technical convenience.

\emph{Remark on the relation to other uses of first-order grammars.}
In this paper the first-order grammars are used for 
slightly different aims than in the works on higher-order grammars
(or higher-order recursion schemes)
and higher-order pushdown automata, where the first order is
a particular case; 
we can exemplify such works 
by~\cite{DBLP:journals/tcs/Courcelle95,DBLP:conf/fossacs/KnapikNU02},
while many other references can be found, e.g., in the survey
papers~\cite{DBLP:conf/lics/Ong15,DBLP:journals/siglog/Walukiewicz16}.
There a grammar is used to describe an infinite labelled tree
(the syntax tree of an infinite applicative term produced by a unique outermost derivation from an initial
nonterminal), and the questions like, e.g., the decidability of monadic
second-order (MSO) properties for such trees are studied.
In this paper, a first-order grammar can be also seen as a tool describing
an infinite tree, namely the tree-unfolding of a nondeterministic
labelled transition system with an initial state. 
The question if this tree
is regular (i.e., if it has only finitely many subtrees)
would correspond to the regularity question studied
in~\cite{DBLP:journals/iandc/Stearns67,DBLP:journals/jacm/Valiant75}; 
but here we ask a different question, namely 
if identifying bisimilar subtrees results in a regular tree.

We can also note that the question if a given first-order grammar
generates a regular tree refers to a particular formalism (namely to the
respective infinite applicative term) while the regularity question
studied here is more ``syntax-independent''.

Some further remarks are given at the end of Section~\ref{sec:prelim}
and in Section~\ref{sec:AddRem}.

\emph{Organization of the paper.}
Section~\ref{sec:prelim} explains the used notions, and states the
result. The proof is then given in
Section~\ref{sec:proof}, and Section~\ref{sec:AddRem} adds a few
additional remarks.
Finally, there is an appendix with some technical constructions that are
not crucial for the proof. 

\emph{Remark.}
A preliminary version of this paper, with a sketch of the proof ideas,
appeared in {Proc.}~MFCS'16.

\section{Basic Notions, and Result}\label{sec:prelim}

In this section we define the basic notions and 
state the result in the form of a theorem.
Some standard definitions are restricted 
when we do not need the full generality.
We finish the section by a note about a transformation of pushdown
automata to
first-order grammars. 

By $\Nat$ and $\Natpos$ we denote the 
sets
of nonnegative integers and of positive integers, respectively.
By $[i,j]$, for $i,j\in\Nat$, we denote the set $\{i,i{+}1,\dots,j\}$.
For a set $\calA$, by $\calA^*$ we denote the set of finite
sequences of elements of $\calA$, which are also called \emph{words}
(over $\calA$).
By $|w|$ we denote the
\emph{length} of 
$w\in \calA^*$, and 
by $\varepsilon$ the \emph{empty sequence};
hence $|\varepsilon|=0$. We put
$\calA^+=\calA^*\smallsetminus\{\varepsilon\}$,
$w^0=\varepsilon$, and $w^{j+1}=ww^j$ for $j\in\Nat$; $w^\omega$
denotes the infinite sequence $www\cdots$.

\subparagraph*{Labelled transition systems.}
A \emph{labelled transition system}, an \emph{LTS} for short,
is a tuple 
$\calL=(\calS,\act,(\gt{a})_{a\in{\act}})$
where $\calS$ is a finite or countable
set of \emph{states},
$\act$ is a finite 
set of \emph{actions} (or \emph{letters}),
and $\gt{a}\subseteq \calS\times\calS$ is a set of
\emph{$a$-transitions} (for each $a\in\act$). 
We say that $\calL$ is 
a \emph{deterministic LTS} if for each pair 
$s\in\calS$, $a\in\act$ there is 
at most one $s'$ such that $s\gt{a}s'$ (which stands for
$(s,s')\in\gt{a}$).
By $s\gt{w}s'$, where 
$w=a_1a_2\dots a_n\in
\act^*$,
we denote 
that there is a \emph{path}
$s=s_0\gt{a_1}s_1\gt{a_2}s_2\cdots\gt{a_n}s_n=s'$;
if $s\gt{w}s'$, then  
$s'$ is \emph{reachable from} $s$. By $s\gt{w}$ we denote that $w$ is
\emph{enabled in} $s$, i.e., $s\gt{w}s'$ for some $s'$.
If $\calL$ is deterministic, then  $s\gt{w}s'$ and $s\gt{w}$ also denote a unique path.
 
\subparagraph*{Bisimilarity.}
Given $\calL=(\calS,\act,(\gt{a})_{a\in\act})$, 
a \emph{set} $\calB\subseteq \calS\times\calS$  
\emph{covers} 
$(s,t)\in  \calS\times\calS$ if 
for each $s\gt{a}s'$ there is $t\gt{a}t'$ such that 
$(s',t')\in \calB$, and for each   $t\gt{a}t'$ there is $s\gt{a}s'$
such that 
$(s',t')\in \calB$.
For $\calB, \calB'\subseteq \calS\times\calS$
we say that $\calB'$ \emph{covers} $\calB$ if $\calB'$
covers each $(s,t)\in \calB$.
A set $\calB\subseteq \calS\times\calS$
is a \emph{bisimulation} if $\calB$ covers $\calB$.
States $s,t\in\calS$ are \emph{bisimilar},
written $s\sim t$, if there is a bisimulation
$\calB$ containing $(s,t)$. 
A standard 
fact is that 
$\sim\,\subseteq \calS\times\calS$ is an equivalence relation,
and it is the largest
bisimulation, namely the union of all bisimulations.

\begin{figure}[ht]
\centering
\includegraphics[scale=0.5]{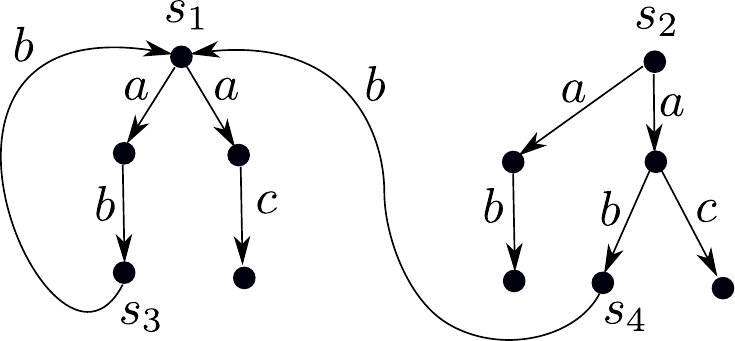}
\caption{Example of a finite (nondeterministic) labelled transition
	system}\label{fig:basicLTS}
\end{figure}

E.g., in the LTS $(\calS,\act,(\gt{a})_{a\in\act})$ in Fig.~\ref{fig:basicLTS} we have $s_3\sim s_4$ and
$s_1\not\sim s_2$ (though $s_1,s_2$ are trace-equivalent, i.e., the
sets $\{w\in\act^*\mid s_1\gt{w}\}$ and  $\{w\in\act^*\mid s_2\gt{w}\}$ are the same).

\subparagraph{Semantic finiteness.} 
Given $\calL=(\calS,\act,(\gt{a})_{a\in\act})$,
we say
that $s_0\in\calS$
is \emph{finite up to bisimilarity},
or \emph{bisim-finite} for short, 
if there is some state $f$ in some
finite LTS such that $s_0\sim f$; otherwise 
$s_0$ is \emph{infinite up to bisimilarity},
or \emph{bisim-infinite}.
We should add that when comparing states from
different LTSs, we implicitly refer to the disjoint union of these
LTSs.

\begin{figure}[ht]
\centering
\includegraphics[scale=0.4]{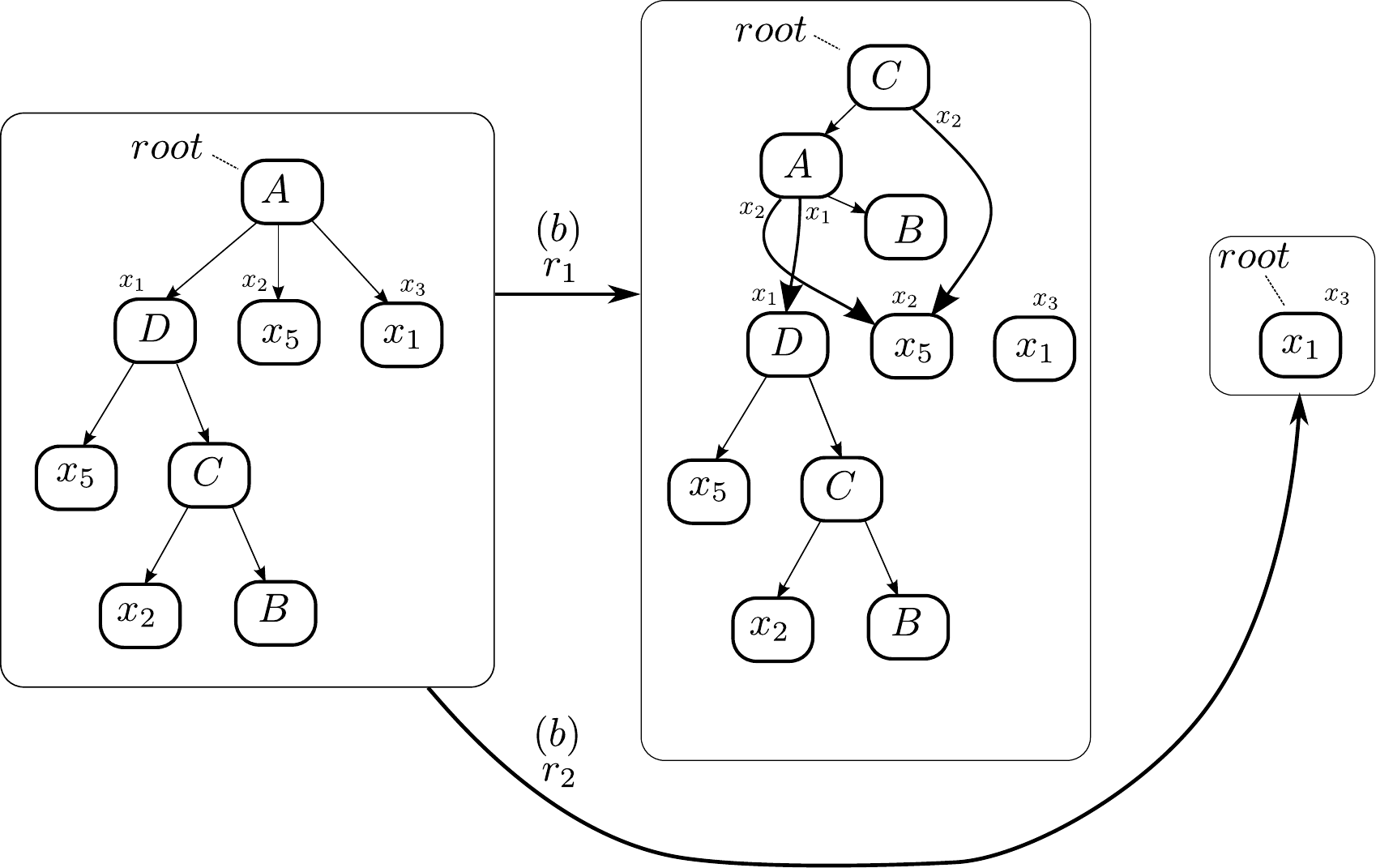}
\caption{One $r_1$-transition and one $r_2$-transition 
in $\calL^{\ltsrul}_{\calG}$ (both are $b$-transitions in
$\calL^{\ltsact}_{\calG}$)}\label{fig:ruleapplic1}
\end{figure}

\subparagraph{First-order terms, regular terms, finite graph presentations.}
We will consider LTSs with countable sets of states in which the states
are
first-order regular terms. 
(In Fig.~\ref{fig:basicLTS} the states are depicted as unstructured
black dots; Fig.~\ref{fig:ruleapplic1} depicts three states of an LTS
with terms ``inside the black dots''.)

The terms are built from \emph{variables}
taken from a fixed countable set
$$\var=\{x_1,x_2,x_3,\dots\}$$ and from 
\emph{function symbols}, also called \emph{(ranked) nonterminals},
from some specified finite set $\calN$; each $A\in\calN$ has 
$\arity(A)\in\Nat$.
We reserve symbols $A,B,C,D$ to range over nonterminals, and 
$E,F,G,H$ 
to range over 
terms.

\begin{figure}[ht]
\centering
\includegraphics[scale=0.4]{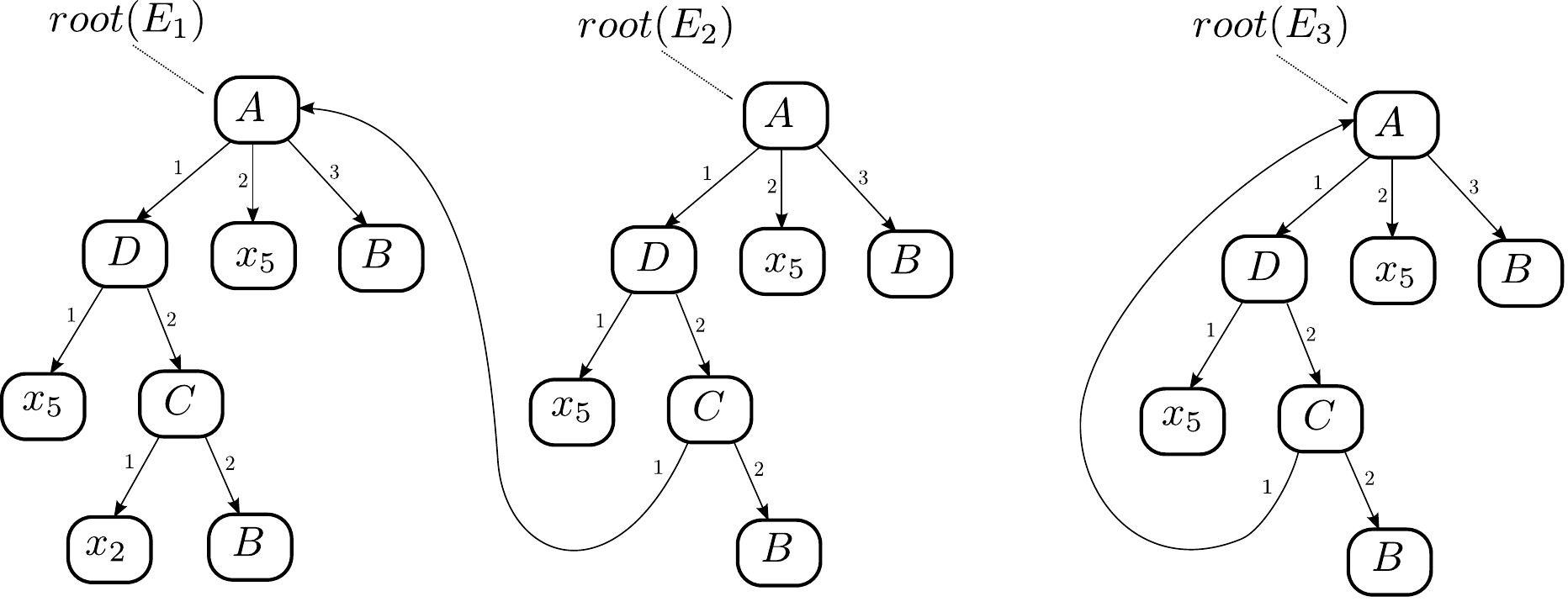}
\caption{Finite terms $E_1$, $E_2$, and 
a graph presenting a regular infinite term $E_3$}\label{fig:basicterm}
\end{figure}

E.g., on the left in Fig.~\ref{fig:basicterm} we can see the syntactic tree
of a term $E_1$, namely of $E_1=A(D(x_5,C(x_2,B)),x_5,B)$,
where the arities of nonterminals $A,B,C,D$ are $3,0,2,2$, respectively.
The numbers at the arcs just highlight the fact that the outgoing arcs 
of each node are ordered.

We identify terms with their syntactic trees.
Thus a \emph{term over}
$\calN$ is (viewed as) a rooted, ordered, finite or infinite tree where each node
has a label from $\calN\cup\var$; if the label of a node is $x_i\in\var$, 
then the node has no successors, and if the label is $A\in\calN$, then 
it has $m$ (immediate) successor-nodes where $m=\arity(A)$.
A subtree of a term 
$E$ is also called a
\emph{subterm} of $E$. We make no difference between isomorphic
(sub)trees, and thus a subterm can have more (maybe infinitely
many) \emph{occurrences} in $E$.  Each \emph{subterm-occurrence} has
its (nesting) \emph{depth in} $E$, which is its (naturally defined) 
distance from the root of $E$.

E.g., $C(x_2,B)$ 
is a subterm of the term $E_1$  in  Fig.~\ref{fig:basicterm},
with one depth-$2$
occurrence.
The term $B$ has two occurrences in $E_1$,
one in depth $1$ and another in depth $3$.

We also use the standard notation for terms: 
we write $E=x$
or $E=A(G_1,\dots, G_m)$ with the obvious meaning; in
the former case we have $x\in\var=\{x_1,x_2,\dots\}$ and
$\termroot(E)=x$, in the
latter case we have $A\in\calN$, $\termroot(E)=A$, $m=\arity(A)$, and
$G_1,\dots,G_m$ are the ordered  depth-$1$ occurrences of  subterms of
$E$, which are also called the \emph{root-successors} in $E$.

A \emph{term} is \emph{finite} if the respective tree is finite. 
A (possibly infinite) \emph{term} is \emph{regular}
if it has only finitely many subterms (though the subterms may be infinite and
can have infinitely many occurrences). 
We note that any regular term has at least one \emph{graph
presentation}, {i.e.}, a finite directed graph with a designated root
where each node has a
label from $\calN\cup\var$; if the label of a node is $x\in\var$,
then the node has no outgoing arcs, and if the label is  $A\in\calN$,
then it has $m$
 ordered outgoing arcs where $m=\arity(A)$.
(A graph presentation of an infinite regular term $E_3$ is on the right in
Fig.~\ref{fig:basicterm}.) 

The standard tree-unfolding of a 
graph presentation is the respective term, which is infinite if there are cycles in
the graph. There is a bijection between 
the nodes in the \emph{least} graph presentation of $E$ 
and (the roots of) the subterms of $E$.
(To get the least graph presentation of $E_3$ in Fig.~\ref{fig:basicterm},
we should unify the roots of the same subterms, here the nodes
labelled with $B$ and the nodes labelled with $x_5$.) 

\textbf{Convention.}
In what follows, by a ``term'' we mean a ``regular term'' unless the
context makes clear that the term is finite.
(We do not consider non-regular terms.)
By $\trees_{\calN}$ we denote the set of all (regular) terms over
a set $\calN$ of (ranked) nonterminals (and over the set $\var$ of
variables).
As already said, we reserve symbols $A,B,C,D$ to range over nonterminals, and 
$E,F,G,H$ 
to range over 
(regular) terms.

\subparagraph*{Substitutions, associative composition, ``iterated''
 substitutions.}\hfill

A \emph{substitution} $\sigma$ is a mapping
$\sigma:\var\rightarrow\trees_{\calN}$ whose 
\emph{support}
\begin{center}
$\support(\sigma)=\{x\in\var\mid \sigma(x)\neq x\}$ 
\end{center}
is \emph{finite};
we reserve the symbol $\sigma$ for substitutions.
By $[x_{i_1}/E_1, x_{i_2}/E_2, \dots, x_{i_k}/E_k]$, where $i_j\neq
i_{j'}$ when $j\neq j'$, we denote the substitution 
$\sigma$ such that $\sigma(x_{i_j})=E_j$ for all $j\in[1,k]$ and
$\sigma(x)=x$ for all
$x\in\var\smallsetminus\{x_{i_1},x_{i_2},\dots,x_{i_k}\}$.

By \emph{applying a substitution} $\sigma$ {to 
a term} $E$ we get the term $E\sigma$ 
that arises from $E$ by replacing each occurrence of $x\in\var$ with
$\sigma(x)$.
(Given graph presentations, 
in the graph of $E$ we just redirect each arc
leading to $x$ towards the root of $\sigma(x)$, which also includes the
special ``root-designating arc'' when $E=x$.) Hence $E=x$ implies
$E\sigma=x\sigma=\sigma(x)$.
(E.g., for the terms in  Fig.~\ref{fig:basicterm} we have
$E_2=E_1[x_2/E_1]$.) 

The natural \emph{composition of substitutions}, where
$\sigma=\sigma_1\sigma_2$ is defined by
$x\sigma=(x\sigma_1)\sigma_2$,
can be easily verified to be
associative. We thus write simply $E\sigma_1\sigma_2$ when meaning 
$(E\sigma_1)\sigma_2$ or  $E(\sigma_1\sigma_2)$. 
We let $\sigma^0$ 
be the empty-support substitution, and we put
$\sigma^{i+1}=\sigma\sigma^i$ for $i\in\Nat$.
We will also use 
the limit substitution
\begin{center}
$\sigma^\omega=\sigma\sigma\sigma\cdots$
\end{center}
when this is well-defined, i.e., 
when there is no ``unguarded cycle''
$x_{i_1}\sigma =x_{i_2}$,
$x_{i_2}\sigma =x_{i_3}$, $\dots$, $x_{i_{k-1}}\sigma =x_{i_k}$,
 $x_{i_{k}}\sigma =x_{i_1}$
where $x_{i_1}\neq x_{i_2}$.
In this case we can formally define $\sigma^\omega$ as the unique
substitution satisfying the following conditions
for each $x\in\var$: if $x\sigma^k\in\var$ for all $k\in\Nat$,
then $x\sigma^\omega=x'$ for $x'\in\var$ where $x\sigma^k=x'$ and 
$x'\sigma=x'$ for some $k$ (such unique $x'$ must exist since 
$\support(\sigma)$ is finite and there is no ``unguarded cycle'');
if there is the least $k\in\Nat$ such that $x\sigma^k=E\not\in\var$
(hence $E$ has a nonterminal root), then 
$x\sigma^\omega=E\sigma^\omega$.
(E.g., in  Fig.~\ref{fig:basicterm} we have
$E_3=E_1\sigma^\omega$ for
$\sigma=[x_2/E_1]$.)
In fact, we will use $\sigma^\omega$ only for
special ``colour-idempotent'' substitutions $\sigma$ defined later.

\begin{figure}[ht]
\centering
\includegraphics[scale=0.4]{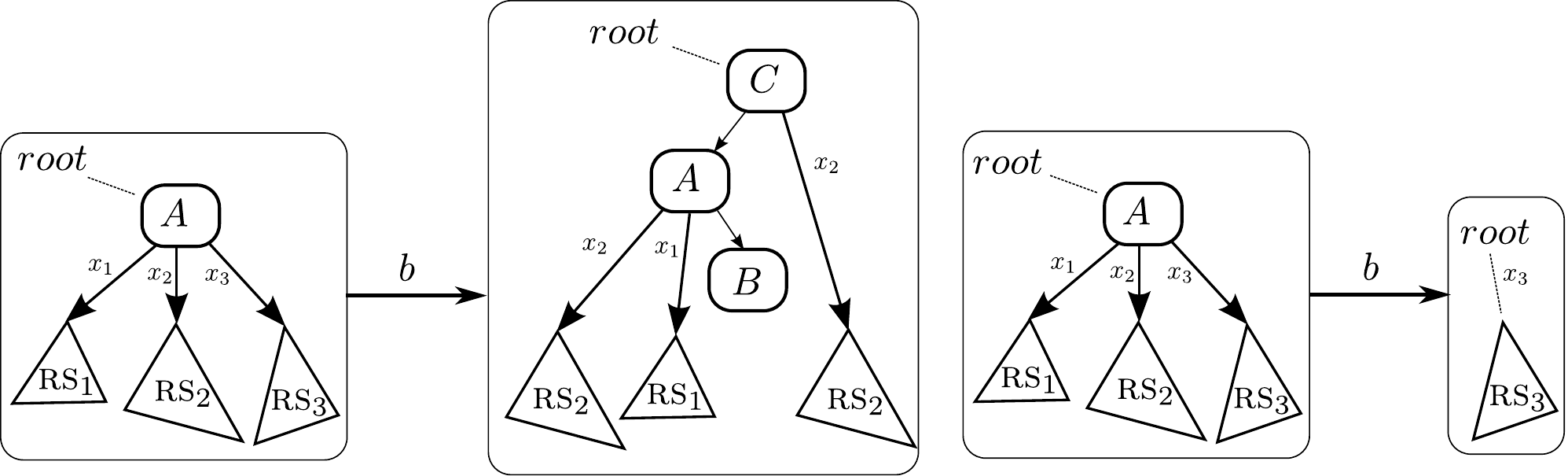}
\caption{Depiction of rules $A(x_1,x_2,x_3)\gt{b}C(A(x_2,x_1,B),x_2)$
and $A(x_1,x_2,x_3)\gt{b}x_3$}\label{fig:basictransition}
\end{figure}

\subparagraph*{First-order grammars.}
The set $\trees_{\calN}$ (of regular terms over a finite set $\calN$ of
nonterminals) will serve us as the set of states of an LTS. 
The transitions will be determined by a finite set of 
(schematic) \emph{root-rewriting} rules, illustrated 
in Fig.~\ref{fig:basictransition}.
This is now defined
formally.

A \emph{first-order grammar}, 
or just a \emph{grammar} for short, is a tuple
$\calG=(\calN,\act,\calR)$ where 
$\calN$
is a finite set of 
ranked \emph{nonterminals} (viewed as function symbols with arities), 
$\act$
is a finite set of \emph{actions} (or letters), 
and 
$\calR$
is
a finite set of 
\emph{rules} of the form
\begin{equation}\label{eq:rewrule}
A(x_1,x_2,\dots, x_m)\gt{a} E
\end{equation}
where $A\in \calN$, $\arity(A)=m$, 
$a\in\act$,
and $E$ is a
finite
term over $\calN$ 
in which each occurring variable 
is
from the set $\{x_1,x_2,\dots,x_m\}$.

\medskip

\emph{Example.}
Fig.~\ref{fig:basictransition} shows 
a 
rule
$A(x_1,x_2,x_3)\gt{b}C(A(x_2,x_1,B),x_2)$, and 
a 
rule 
$A(x_1,x_2,x_3)\gt{b}x_3$. The depiction stresses
that the variables $x_1,x_2,x_3$ serve as
the ``place-holders'' for the root-successors ($\textsc{rs}$), 
{i.e.}~the depth-1 occurrences of subterms of a term with the
root $A$; the (root of the) term might be
rewritten by performing action $b$ (as defined below).

\begin{figure}[ht]
\centering
\includegraphics[scale=0.4]{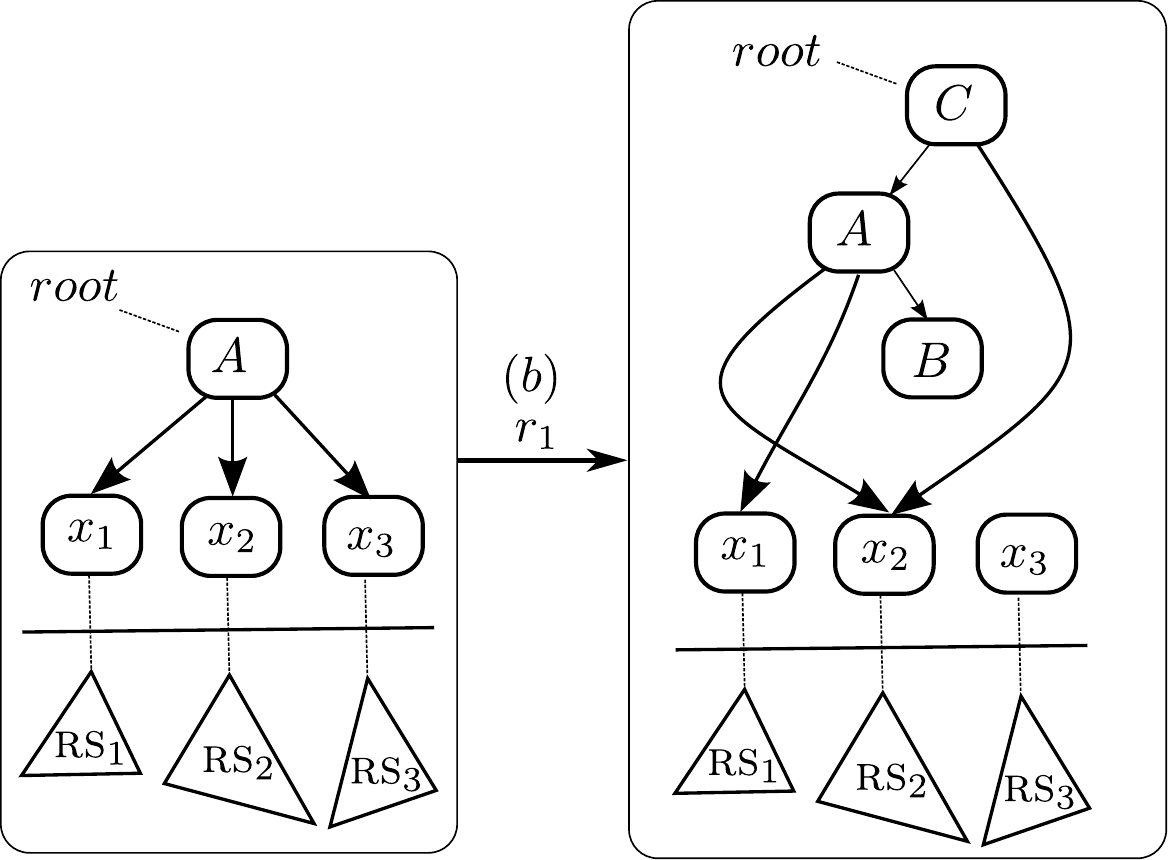}
\caption{Another presentation of the rule
$r_1:A(x_1,x_2,x_3)\gt{b}C(A(x_2,x_1,B),x_2)$}\label{fig:basictransition2}
\end{figure}

\subparagraph{LTSs generated by grammars.}
Given $\calG=(\calN,\act,\calR)$, 
by $\calL^{\ltsrul}_{\calG}$ we denote the (\emph{rule-based}) LTS
$\calL^{\ltsrul}_{\calG}=(\trees_{\calN},\calR,(\gt{r})_{r\in\calR})$
where each rule $r$ of the form
$A(x_1,x_2,\dots, x_m)\gt{a} E$ 
induces
transitions $A(x_1,\dots, x_m)\sigma\gt{r}E\sigma$
for all substitutions $\sigma$. (In fact, only the restrictions of
substitutions $\sigma$ to the domain $\{x_1,\dots,x_m\}$ matter.)
The transition induced by  $\sigma$
with $\support(\sigma)=\emptyset$ is
$A(x_1,\dots, x_m)\gt{r}E$.

\medskip

\begin{figure}[ht]
\centering
\includegraphics[scale=0.38]{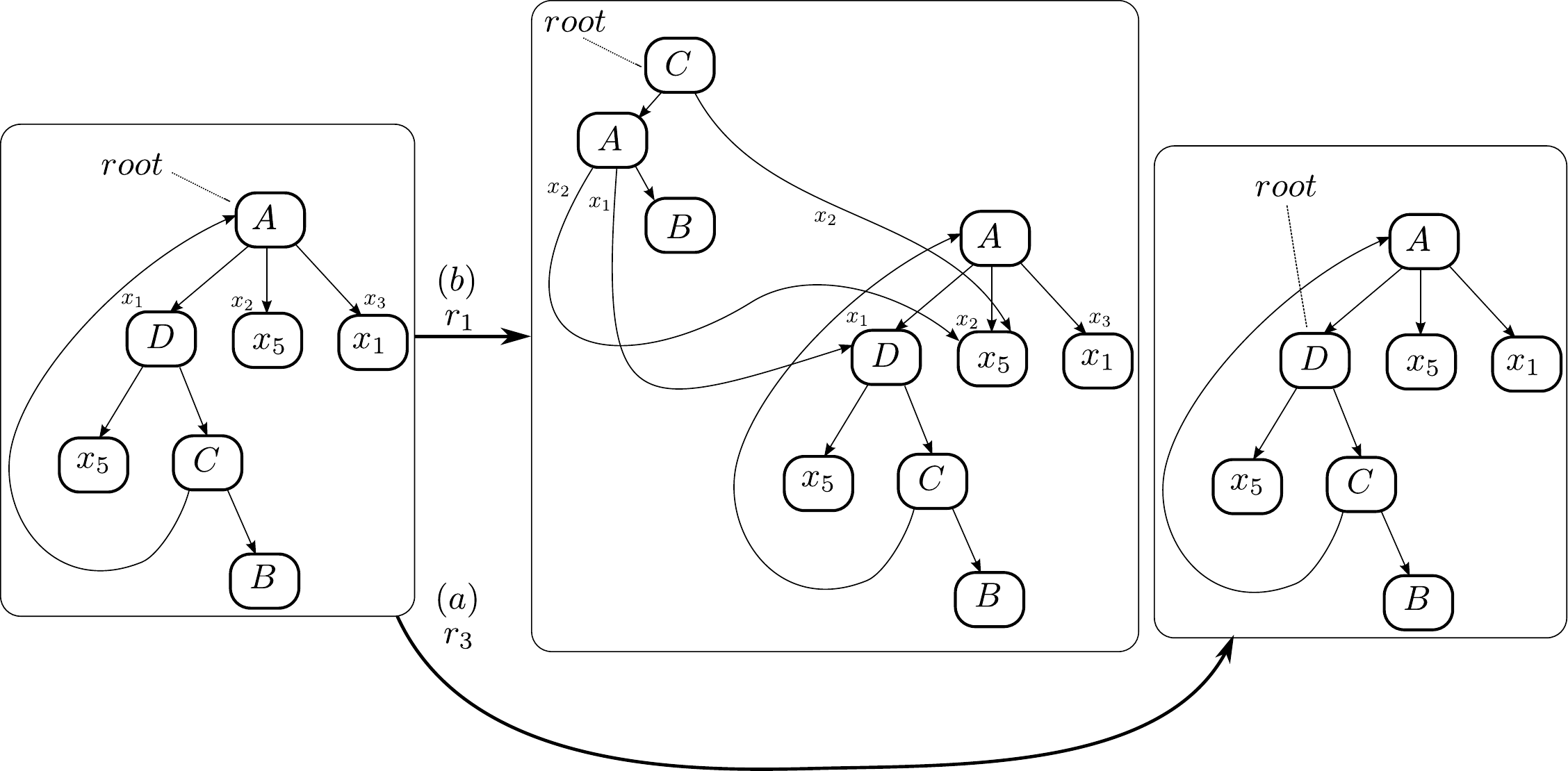}
\caption{Applying 
$r_1:A(x_1,x_2,x_3)\gt{b}C(A(x_2,x_1,B),x_2)$ and 
$r_3:A(x_1,x_2,x_3)\gt{a}x_1$
to a graph of an (infinite regular) term}\label{fig:ruleapplic2}
\end{figure}

\emph{Example.}
To continue the example from  Fig.~\ref{fig:basictransition},
in
Fig.~\ref{fig:basictransition2} we can see another 
 depiction of the rule 
 $A(x_1,x_2,x_3)\gt{b}C(A(x_2,x_1,B),x_2)$, 
 denoted $r_1$.
This makes more explicit that the application of the same substitution
to both sides yields a transition; 
it also highlights the fact
that  $\textsc{rs}_3$ ``disappears'' by applying the rule
since it loses the connection
with the root. 
	Fig.~\ref{fig:ruleapplic1} shows two examples 
	of transitions in $\calL^{\ltsrul}_{\calG}$.
One is 	generated by the 
rule $r_1$ of the
form $A(x_1,x_2,x_3)\gt{b}C(A(x_2,x_1,B),x_2)$
(depicted in Figures~\ref{fig:basictransition}
and~\ref{fig:basictransition2}),
	and the other by  the 
	rule $r_2$ of the
	form $A(x_1,x_2,x_3)\gt{b}x_3$; in both cases we apply
	the substitution $\sigma=[x_1/D(x_5,C(x_2,B)), x_2/x_5,
x_3/x_1]$ to (the both sides of) the respective rule.
	The small symbols $x_1,x_2,x_3$ in Fig.~\ref{fig:ruleapplic1}
	are only auxiliary, 
	highlighting the use of our rules, and they are no part of the
	respective terms. The middle term is here given by 
	an (acyclic) graph presentation of its syntactic tree (and the
	node with label $x_1$ is no part of it).
	Fig.~\ref{fig:ruleapplic2} shows the transitions
	resulting by the applications of two rules 
	to a graph presenting an infinite regular term.
	(The small symbols $x_1,x_2,x_3$ are again just auxiliary.)
\medskip

\emph{Remark.}
Since the rhs (right-hand sides) $E$ in the rules~(\ref{eq:rewrule}) 
are finite, all
terms reachable from a finite term are finite.
It is technically convenient to have the rhs finite 
while including regular terms into our LTSs.
We just remark that this is not crucial, since 
the ``regular rhs'' version can be easily ``mimicked'' 
by the ``finite rhs'' version.

\medskip

By definition the LTS $\calL^{\ltsrul}_{\calG}$ is deterministic
(for each $F$ and $r$ there is at most one $H$ such that $F\gt{r}H$).
We note that \emph{variables} are \emph{dead} (have no outgoing
transitions).
We also note  
that $F\gt{w}H$ implies that
each variable occurring in $H$ also occurs in $F$ (but not necessarily
vice versa).

The deterministic rule-based LTS $\calL^{\ltsrul}_{\calG}$ is helpful
technically,
but we are primarily interested in the (generally nondeterministic)
\emph{action-based} LTS 
$\calL^{\ltsact}_{\calG}=(\trees_{\calN},\act,(\gt{a})_{a\in\act})$
where each rule $A(x_1,\dots, x_m)\gt{a}E$
induces  the transitions
$A(x_1,\dots, x_m)\sigma\gt{a}E\sigma$ for all substitutions $\sigma$.
(Figures~\ref{fig:ruleapplic1} and~\ref{fig:ruleapplic2} also
	show transitions in $\calL^{\ltsact}_{\calG}$, when we ignore
	the symbols $r_1,r_2,r_3$ and consider the ``labels'' $b,a$
	instead. Figure~\ref{fig:ruleapplic1} also exemplifies
	nondeterminism in $\calL^{\ltsact}_{\calG}$, since there are
	two different outgoing $b$-transitions from a state.)

Given a grammar $\calG=(\calN,\act,\calR)$, two \emph{terms} from
$\trees_{\calN}$ are
\emph{bisimilar} if they are bisimilar as states in  the action-based LTS
$\calL^{\ltsact}_{\calG}$.
By our definitions all variables are bisimilar, since they are dead terms.
The variables serve us primarily as ``place-holders for
subterm-occurrences'' in
terms (which might themselves be variable-free);
such a use of variables as place-holders has been 
already exemplified  
in the rules~(\ref{eq:rewrule}).

\subparagraph*{Main result, and its relation to pushdown automata.}

We now state the theorem, to be proven in the next section, and we
mention
why the result also applies to pushdown automata (PDA) with deterministic popping
$\varepsilon$-steps.

\begin{theorem}\label{th:decidsemfinit}
There is an algorithm that, given a  grammar
$\calG=(\calN,\act,\calR)$ and (a finite graph presentation of) a term
$E_0\in\trees(\calN)$, decides if $E_0$ is bisim-finite
(i.e., if $E_0\sim f$ for a state $f$ in some finite LTS).
\end{theorem}

A transformation of
(nondeterministic) PDA in which deterministic popping $\varepsilon$-steps
are allowed to first-order grammars (with no $\varepsilon$-steps) is
recalled 
in 
the appendix.
It makes clear that the semantic finiteness ({w.r.t.}~bisimilarity) of PDA with 
deterministic popping $\varepsilon$-steps is
also decidable. In fact, the problems are interreducible;
the close relationship between (D)PDA and first-order schemes
has been long known~(see, e.g.,~\cite{CourcelleHandbook}).
The proof of Theorem~\ref{th:decidsemfinit} presented here
uses the fact that bisimilarity of first-order grammars is decidable;
this was shown for the above mentioned PDA model by
S\'enizergues~\cite{Seni05}, and a direct proof in the
first-order-term framework
was presented in~\cite{DBLP:conf/icalp/Jancar14}.

We note that for PDA where popping  $\varepsilon$-steps can be in conflict
with ``visible'' steps bisimilarity is
already undecidable~\cite{DBLP:journals/jacm/JancarS08}; hence 
the proof
presented here does not yield 
the decidability of semantic finiteness in this more general model.
The decidability status of semantic finiteness is also unclear
for \emph{second-order} PDA (that operate on a stack of stacks; besides the
standard work on the topmost stack, they can also 
push a copy of the topmost stack or to pop the topmost stack in one
move).
Bisimilarity is undecidable for second-order PDA even without any use
of $\varepsilon$-steps~\cite{DBLP:conf/fsttcs/BroadbentG12} (some
remarks are also added in~\cite{DBLP:journals/corr/abs-1303-0780}).

\section{Proof of Theorem~\ref{th:decidsemfinit}}\label{sec:proof}

\subsection{Computability of eq-levels, and semidecidability of
bisim-finiteness}

We will soon note that the
semidecidability of bisim-finiteness is clear, but we first
 recall the computability of eq-levels, which is one crucial
 ingredient in our proof of semidecidability of bisim-infiniteness.

\subparagraph*{Stratified equivalence, and eq-levels.}
 
 Assuming an LTS $\calL=(\calS,\act,(\gt{a})_{a\in\act})$, we
put $\sim_0=\calS\times\calS$, and define 
$\sim_{k+1}\subseteq\calS\times\calS$
(for $k\in\Nat$)
as the set of pairs 
covered by $\sim_{k}$. 
(Hence $s\sim_{k+1}t$ iff
for each $s\gt{a}s'$ there is $t\gt{a}t'$ such that 
$s'\sim_k t'$ and for each   $t\gt{a}t'$ there is $s\gt{a}s'$
such that $s'\sim_k t'$.)

We easily verify that $\sim_k$ are equivalence relations, and
that $\sim_0\,\supseteq\,
\sim_{1}\,\supseteq\,\sim_2\,\supseteq\,\cdots\cdots\supseteq \sim$.
For the (first infinite) ordinal $\omega$ we put 
$s\sim_\omega t$ if $s\sim_k t$ for all $k\in\Nat$; hence 
$\sim_\omega=\bigcap_{k\in\Nat}\sim_k$.
We do not need to consider ordinals bigger than $\omega$, 
due to the following restriction.
An LTS $\calL=(S,\act,(\gt{a})_{a\in\act})$ is \emph{image-finite} if 
 the set $\{s'\in S\mid s\gt{a}s'\}$ is finite for each 
 pair $s\in\calS$, $a\in\act$.  Our grammar-generated LTSs
 $\calL^{\ltsact}_\calG$ are obviously image-finite
 (while  $\calL^{\ltsrul}_\calG$ are even deterministic). We thus 
 further restrict ourselves to image-finite LTSs. In fact, since we consider LTSs with
 finite sets of actions, we are thus even restricted to \emph{finitely
 branching} LTSs, where the set $\{s'\in S\mid s\gt{a}s'$ for some
 $a\in\Sigma\}$ is
 finite for each $s\in S$.

It is a standard fact that 
\[\sim=\sim_\omega=\bigcap_{k\in\Nat}\sim_k\]
in image-finite LTSs
(as also mentioned, e.g., in~\cite{Milner1989}); 
indeed, it is straightforward to check that in such an LTS the set
$\bigcap_{k\in\Nat}\sim_k$
is covered by itself, hence $\sim_\omega$ is a bisimulation (which
entails
$\sim_\omega\subseteq\sim$, and thus $\sim_\omega=\sim$).

Given a (finitely branching) LTS
$\calL=(\calS,\act,(\gt{a})_{a\in\act})$, to each (unordered) pair $s,t$
of states 
we attach their \emph{equivalence level} (eq-level): 
\begin{center}
$\eqlevel(s,t)=\max\,\{k\in\Nat\cup\{\omega\}\mid s\sim_k t\}$.
\end{center}
(In Fig.~\ref{fig:basicLTS} we have $\eqlevel(s_3,s_4)=\omega$,
$\eqlevel(s_1,s_3)=\eqlevel(s_1,s_4)=0$,
$\eqlevel(s_1,s_2)=1$.)
It is useful to observe:
\begin{proposition}\label{prop:easyeqlevel}
If $s\sim s'$ then $\eqlevel(s,t)=\eqlevel(s',t)$,
for all states $s,s',t$.
\end{proposition}

\begin{proof}
Suppose $s\sim s'$, i.e., $s\sim_k s'$ for all $k\in\Nat$.
For each $k\in\Nat$ we then have that $s\sim_k t$ implies $s'\sim_k t$ and
$s\not\sim_{k} t$ implies $s'\not\sim_{k} t$, since 
$\sim_k$ are equivalence relations. 
\end{proof}

\subparagraph{Eq-levels are computable for first-order grammars.}
We now recall a variant of the fundamental decidability theorem
shown by S\'enizergues in~\cite{Seni05}; it
will be used as an important ingredient in the 
proof of Theorem~\ref{th:decidsemfinit}.

\begin{theorem}\label{thm:pdabisimdecid}\textnormal{\cite{Seni05}}
There is an algorithm that, given $\calG=(\calN,\act,\calR)$ and
$E_0,F_0\in\trees(\calN)$,
computes $\eqlevel(E_0,F_0)$ in $\calL^{\ltsact}_{\calG}$
(and thus also decides if $E_0\sim F_0$).
\end{theorem}

The crucial thing is that we can decide if $E_0\sim F_0$
(by the algorithm from~\cite{Seni05}, or by the alternative algorithm
presented in~\cite{DBLP:conf/icalp/Jancar14} directly in the framework
of first-order grammars).
If $E_0\not\sim F_0$, then a straightforward
brute-force algorithm finds the least $k{+}1\in\Nat$ such that 
$E_0\not\sim_{k+1} F_0$, thus finding that $\eqlevel(E_0,F_0)=k$.

\subparagraph*{Semidecidability of bisim-finiteness.} 
Given $\calG$ and $E_0$,
we can systematically generate all finite LTSs, 
presenting them by first-order grammars with nullary
nonterminals (which then coincide with states); for each state $f$ of
each generated system we can check if $E_0\sim f$ 
by Theorem~\ref{thm:pdabisimdecid}.
In fact, Theorem~\ref{thm:pdabisimdecid} is not crucial here, 
since the decidability of $E_0\sim f$ 
can be shown in a much simpler way (see, e.g.,~\cite{Kucera10}).

\subsection{Semidecidability of bisim-infiniteness} 

In Section~\ref{subsec:witnessfacts} we note a few simple general
facts on bisim-infiniteness, and also note the obvious compositionality    
(congruence properties) of bisimulation equivalence in our
framework of first-order terms.

In Section~\ref{subsec:simplewitnesses} we describe some finite
structures  that are candidates for witnessing 
bisim-infiniteness of a given term $E_0$;
such a candidate is, in fact, a rule sequence $uw$ such that
the infinite (ultimately periodic) word $uw^\omega$ is performable from
$E_0$ in $\calL^{\ltsrul}_\calG$.
Then we show an algorithm checking if a
candidate is indeed a witness, i.e., if the respective infinite path
$E_0\trans{u}\trans{w}\trans{w}\trans{w}\cdots$ 
visits terms from infinitely many equivalence
 classes. The crucial idea is that we can naturally
define a (regular) term, called the \emph{limit} $\LIMIT=E_\omega$,
that could be viewed as
``reached'' from $E_0$ by performing the infinite word $uw^{\omega}$. The
terms $E_j$ such that $E_0\trans{u}\trans{w^j}E_j$ will approach
$E_\omega$ syntactically with increasing $j$ ($E_j$ coincides
with $E_\omega$ up to depth $j$ at least), which also entails that 
$\eqlevel(E_j,E_\omega)$ will grow above any bound. 
If we can verify that $\eqlevel(E_j,E_\omega)$ are finite for
infinitely many $j$, in particular if $\eqlevel(E_j,E_\omega)$ never
reaches $\omega$ (hence $E_j\not\sim E_\omega$ for all $j$),
then $uw$ is indeed a witness of bisim-infiniteness of $E_0$;
Theorem~\ref{thm:pdabisimdecid} 
will play an important role in such a verification.

In  Section~\ref{subsec:infhaswitness} we  show that each
bisim-infinite term has a witness of the above form. 
Here we will also use the infinite Ramsey theorem for a technical
simplification. 
By this a proof 
of Theorem~\ref{th:decidsemfinit} will be finished.

\subsubsection{Some general facts on bisim-infiniteness, 
and compositionality of terms}\label{subsec:witnessfacts}

\subparagraph{Bisimilarity quotient.}
Given an LTS $\calL=(\calS,\act,(\gt{a})_{a\in\act})$,
the 
\emph{quotient-LTS} $\calL_\sim$
is the tuple
$(\{\,[s]_\sim\mid s\in \calS\,\},\act,(\gt{a})_{a\in\act})$ where 
$[s]_\sim=\{s'\mid s'\sim s\}$, 
and $[s]_\sim\gt{a}[t]_\sim$ if $s'\gt{a}t'$ for some $s'\in[s]_\sim$
and $t'\in[t]_\sim$; in fact, $[s]_\sim\gt{a}[t]_\sim$ implies 
that for each $s'\in [s]_\sim$ there is $t'\in[t]_\sim$
such that $s'\gt{a}t'$.
We have $s\sim [s]_\sim$, since $\{(s,[s]_\sim)\mid s\in \calS\}$
is a bisimulation
(in the union of $\calL$ and $\calL_\sim$).
We refer to the states of $\calL_\sim$ as to the
\emph{bisim-classes} (of $\calL$).

\subparagraph{A sufficient condition for bisim-infiniteness.}
We recall
that $s_0\in\calS$
is bisim-finite
if there is some state $f$ in a 
finite LTS such that $s_0\sim f$; otherwise 
$s_0$ is 
bisim-infinite.
We observe that $s_0$ is bisim-infinite in $\calL$
iff the
reachability set of $[s_0]_\sim$ in $\calL_\sim$,
{i.e.}~the set of states reachable from $[s_0]_\sim$ in $\calL_\sim$,
is infinite.

We also recall our restriction to finitely branching LTSs
($\{s'\mid s\gt{a}s'$ for some $a\}$ is finite for each $s$), and note
the following:

\begin{proposition}\label{prop:infinbisimchange}
A state $s_0$ of a finitely branching LTS $\calL=(\calS,\act,(\gt{a})_{a\in\act})$
is bisim-infinite iff 
there is an infinite path
$s_0\gt{a_1}s_1\gt{a_2}s_2\gt{a_3}\cdots$  where $s_i\not\sim s_j$ 
for all $i\neq j$.
\end{proposition}

\begin{proof}
	The ``if'' direction is trivial; our goal is thus to show the
	``only if'' direction.
Let $s_0$ be a bisim-infinite state in 
a finitely branching LTS $\calL$.
In the quotient-LTS $\calL_\sim$ we consider the set $\calP$ of all
finite paths 
${C}_0\gt{a_1}C_1\gt{a_2}C_2\cdots\gt{a_k}C_k$ where $C_0=[s_0]_\sim$
and $C_i\neq C_j$ for all $i,j\in[0,k]$, $i\neq j$.
We present $\calP$ as a tree:
the paths in $\calP$ 
are the nodes, the trivial
path $C_0$ being the root, and each 
node ${C}_0\gt{a_1}C_1\gt{a_2}C_2\cdots\gt{a_{k+1}}C_{k+1}$
is a child of the node ${C}_0\gt{a_1}C_1\gt{a_2}C_2\cdots\gt{a_k}C_k$.
This tree is finitely branching (each node has a finite set of
children). If all branches were finite
(in which case each leaf ${C}_0\gt{a_1}C_1\gt{a_2}C_2\cdots\gt{a_k}C_k$
would satisfy that $C_k\gt{a}C$ implies $C=C_i$ for some $i\in[0,k]$),
then the tree would be
finite (by K\"onig's lemma), and thus the set of states in $\calL_\sim$
that are reachable from $[s_0]_\sim$ would be also finite; this would
contradict with the assumption that $s_0$ is bisim-infinite. 
Hence there is an infinite path
${C}_0\gt{a_1}C_1\gt{a_2}C_2\gt{a_3}\cdots$ in $\calL_\sim$  where $C_0=[s_0]_\sim$
and $C_i\neq C_j$ for all $i\neq j$.
Since $s_0\sim C_0$ (in the union of $\calL$ and $\calL_\sim$),
there
must be 
a path 
$s_0\gt{a_1}s_1\gt{a_2}s_2\gt{a_3}\cdots$ in $\calL$
where $s_i\sim C_i$, i.e. $[s_i]_\sim=C_i$, for all $i\in \Nat$; 
for $i\neq j$ we thus have $[s_i]_\sim\neq [s_j]_\sim$, i.e., 
$s_i\not\sim s_j$.
\end{proof}

To demonstrate that $s_0$ is bisim-infinite, it suffices to show that
its reachability set contains states with arbitrarily large
\emph{finite} eq-levels {w.r.t.}~a ``test state'' $t$; we now formalize this observation.

\begin{proposition}\label{prop:increaseqlevel}
Given $\calL=(\calS,\act,(\gt{a})_{a\in\act})$ and states $s_0,t$, if
for every
$e\in\Nat$ there is $s'$ that is reachable from $s_0$
and satisfies $e<\eqlevel(s',t)<\omega$,
then $s_0$ is bisim-infinite.
\end{proposition}

\begin{proof}
If $s_0,t$ satisfy the assumption, then there are $e_1<e_2<e_3<\cdots$
and states $s_1,s_2,s_3,\ldots$ reachable from $s_0$ such that 
$\eqlevel(s_i,t)=e_i$ for all $i\in\Natpos$.
By Proposition~\ref{prop:easyeqlevel} we thus get that $s_i\not\sim
s_j$ for all $i,j\in\Natpos$, $i\neq j$.
Hence from $s_0$ we
can reach states from infinitely many bisim-classes, which entails
that $s_0$ is bisim-infinite.
\end{proof}

\begin{figure}[ht]
\centering
\includegraphics[scale=0.5]{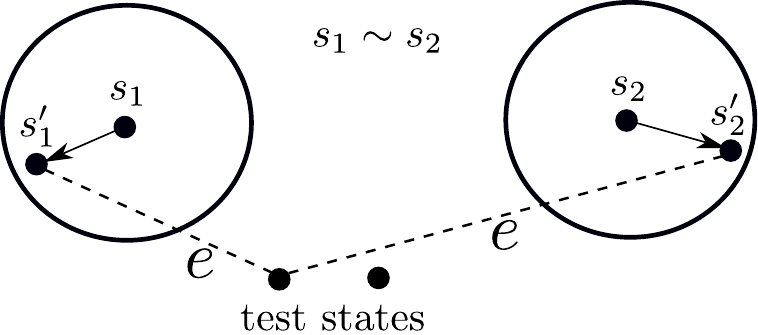}
\caption{Bounded regions of 
bisimilar states yield the same eq-levels {w.r.t.}~test states}\label{fig:regionEL}
\end{figure}

\subparagraph{Eq-levels yielded by states in a bounded region and test
states.}
Our final general observation (tailored to a later use) 
is also straightforward: if two states of an LTS are bisimilar, then the states
in their equally bounded reachability regions must yield the same 
 eq-levels when compared with states from a fixed (test) set.
 This observation is informally depicted in Fig.~\ref{fig:regionEL},
 and formalized in what follows.
 (Despite the depiction in Fig.~\ref{fig:regionEL}, the test states
 can be also inside the regions.)

Given $\calL=(\calS,\act,(\gt{a})_{a\in\act})$,
for any $s\in\calS$ and $d\in\Nat$  (a distance, or a ``radius'')
we put
\begin{equation}\label{eq:defregion}
	\textnormal{
	$\region(s,d)=\{s'\mid  s\gt{w}s'$ 
for some $w\in\act^*$
where $|w|\leq d\}$.
}
\end{equation}
For any $s\in\calS$, $d\in\Nat$, and  $\calT\subseteq\calS$ (a test set),
we define the following subset of $\Nat$ (\emph{finite} TestEqLevels):
\begin{equation}\label{eq:defTEL}
\textnormal{
$\EqLevels(s,d,\calT)=\{e\in\Nat\mid 
e=\eqlevel(s',t)$ for some $s'\in \region(s,d)$ and $t\in \calT \}$.
}
\end{equation}
For $X\subseteq \Nat$, by the supremum $\sup(X)$ we mean
$-1$ if $X=\emptyset$, $\max(X)$ if $X$ is finite and nonempty, and
$\omega$ if $X$ is infinite. 
(The next proposition will be later applied to
the LTSs $\calL^{\ltsact}_\calG$ with finite test sets, hence 
the sets $\region(s,d)$ and $\EqLevels(s,d,\calT)$ will be finite.)

\begin{proposition}\label{prop:regiontest}
If $s_1\sim s_2$, then
$\EqLevels(s_1,d,\calT)=\EqLevels(s_2,d,\calT)$
for all $d\in\Nat$ and $\calT\subseteq\calS$,
which also entails
that
$\sup(\EqLevels(s_1,d,\calT))=\sup(\EqLevels(s_2,d,\calT))$.
\end{proposition}

\begin{proof}
	(Recall Figure~\ref{fig:regionEL}.)
	Suppose $s_1\sim s_2$ and $s'_1\in \region(s_1,d)$;
let $s_1\gt{w}s'_1$ where $|w|\leq d$.
From the definition
of bisimilarity we deduce that 
$s_2\gt{w}s'_2$ for some $s'_2$ such that $s'_1\sim s'_2$; we have 
 $s'_2\in \region(s_2,d)$.
Since $s'_1\sim s'_2$ implies
$\eqlevel(s'_1,t)=\eqlevel(s'_2,t)$ for every $t$
(by Proposition~\ref{prop:easyeqlevel}), the claim is clear. 
\end{proof}

\emph{Remark.}
The fact that $s_1\sim s_2$ and $s_1\gt{a}s'_1$ implies that there is
$s'_2$ such that  $s_2\gt{a}s'_2$ and  $s'_1\sim s'_2$ is a crucial
property of bisimilarity that we use for our decision procedure.
Hence our approach does not apply to trace equivalence or simulation
equivalence. (E.g., the states $s_1,s_2$ in Fig.~\ref{fig:basicLTS}
are trace equivalent but their $a$-successors are from pairwise
different trace equivalence classes.) 
On the other hand, the below mentioned compositionality of
bisimilarity holds 
for other equivalences as well.

\begin{figure}[ht]
\centering
\includegraphics[scale=0.4]{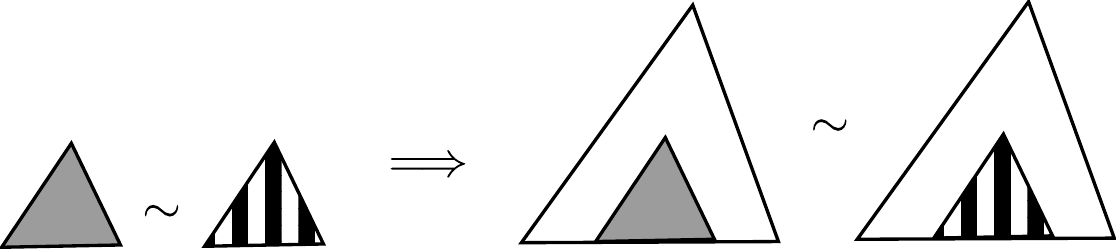}
\caption{Replacing a subterm with an equivalent term does not change
the bisim-class}\label{fig:composit}
\end{figure}

\subparagraph*{Compositionality of states in grammar-generated 
LTSs.}
We assume a grammar $\calG=(\calN,\act,\calR)$, generating the LTS
$\calL^\ltsact_\calG=(\trees_\calN,\act,(\gt{a})_{a\in\act})$ where $\sim$ is bisimulation equivalence on
$\trees_\calN$.
Regarding the congruence properties, in principle it suffices for us to
observe the fact depicted in Fig.~\ref{fig:composit}:
if in a term $E$ we replace a subterm $F$ with $F'$
such that $F'\sim F$ then the resulting term $E'$
satisfies $E'\sim E$.

Hence we also have that
$A(G_1,\dots,G_m)\not\sim A(G'_1,\dots,G'_m)$ implies 
$G_i\not\sim G'_i$ for some $i\in[1,m]$.
Formally, we put $\sigma\sim \sigma'$
if $x\sigma\sim x\sigma'$ for each $x\in\var$, and we note:

\begin{proposition}\label{prop:congruence}
If $\sigma\sim \sigma'$, 
then $E\sigma\sim E\sigma'$.
\\
(Hence $E\sigma\not\sim E\sigma'$ implies 
that $x\sigma\not\sim x\sigma'$ for some $x$ occurring in $E$.)
\end{proposition}	

\begin{proof}
	We show that the set 
$\calB\mathop{=}\sim\mathop{\cup}\,\{(E\sigma,E\sigma')\mid
	E\in\trees_\calN, \sigma\sim\sigma'\}$ is a bisimulation
	(hence $\calB=\,\sim$).
It suffices to consider 
a pair $(E\sigma,E\sigma')$ where $\sigma\sim\sigma'$ and show that it
is covered by $\calB$. If $E=x\in\var$, then  
$(E\sigma,E\sigma')=(x\sigma,x\sigma')$, which entails
$E\sigma\sim E\sigma'$ and $(E\sigma,E\sigma')$ is thus covered
by
the bisimulation $\sim$ (included in $\calB$). If
$\termroot(E)\in\calN$, then each transition $E\sigma\gt{a}G$ can be
written as $E\sigma\gt{a}E'\sigma$ where 
$E\gt{a}E'$; it has the matching transition $E\sigma'\gt{a}E'\sigma'$,
and we have $(E'\sigma,E'\sigma')\in\calB$.
\end{proof}

\subparagraph{Conventions.}
\begin{itemize}
	\item		
To make some later discussions easier, 
we further consider only the \emph{normalized} grammars
$\calG=(\calN,\act,\calR)$,
{i.e.}~those satisfying the following
condition: for each $A\in\calN$ and each $i\in[1,m]$ where
$m=\arity(A)$
there is a
word $w_{(A,i)}\in\calR^+$ such that
$A(x_1,\dots,x_m)\trans{w_{(A,i)}}x_i$ (in $\calL^{\ltsrul}_\calG$).
Hence for each $E\in\trees_\calN$ it is possible to ``sink'' to each of its
subterm-occurrences 
by applying a sequence of the grammar-rules. 
(E.g., in the middle term in Fig.~\ref{fig:ruleapplic2} we can 
``sink'' to  the root-successor term
with the root $A$ by applying some rule sequence $u_1$,
then ``to $D$'' by applying some $u_2$,
then to $C$ by some $u_3$, then to $A$ by some $u_4$, ...)

Such a normalization can be efficiently achieved 
by harmless modifications of the nonterminal arities and of the rules
in $\calR$, while the bisimilarity 
		quotient of the LTS $\calL^{\ltsact}_\calG$ remains the same 
		(up to isomorphism). Now we simply assume this, the details are given 
 in the appendix.
\item
In our notation we use $m$ as the arity of all
nonterminals in the considered grammar, though $m$ is deemed to denote the
\emph{maximum}
arity, in fact. Formally we could replace our expressions
of the form $A(G_1,\dots,G_m)$ 
with $A(G_1,\dots,G_{m_A})$ where $m_A=\arity(A)$, and adjust the
respective discussions accordingly, but it would be unnecessarily
cumbersome.

In fact, such uniformity of arities can be even achieved by a construction
while keeping the previously discussed normalization condition, when a slight
problem with arity $0$ is handled.
The details are also given 
in the appendix.
 
\item
For technical convenience we further 
 view the expressions like $G\gt{w}H$ as referring to 
the deterministic LTS $\calL^{\ltsrul}_\calG$ (hence $w\in\calR^*$
and any expression $G\gt{w}$ refers to a unique path in 
$\calL^{\ltsrul}_\calG$),
while $\sim_k$, $\sim$, 
and the eq-levels are always considered {w.r.t.}~the action-based LTS
$\calL^{\ltsact}_\calG$.
\end{itemize}

\subsubsection{Witnesses of
bisim-infiniteness}\label{subsec:simplewitnesses}

Assuming
a grammar $\calG=(\calN,\act,\calR)$,
we now describe candidates for witnesses of bisim-infiniteness of
terms.
A witness of bisim-infiniteness of $E_0$ will 
be a pair $(u,w)$, $u\in\calR^*$ and $w\in\calR^+$, for which there is
the infinite path $E_0\trans{u}\trans{w}\trans{w}\trans{w}\cdots$ and
it
visits infinitely many bisim-classes. We first put a technically convenient
restriction on the considered ``iterated'' words $w$, and then define
candidates for witnesses formally.

\subparagraph*{Stairs, stair substitutions,  colour-idempotent
substitutions and stairs.}
 A~nonempty sequence $w=r_1r_2\dots
r_\ell\in\calR^+$ of rules is a \emph{stair} 
if we have $A(x_1,\dots,x_m)\gt{w}F$
where $A(x_1,\dots,x_m)$ is the left-hand side of 
the rule $r_1$ and
$\termroot(F)\in\calN$ (i.e., $F\not\in\var$).

E.g., for the rules $r_1:A(x_1,x_2)\gt{a}C(C(x_2,B(x_2,x_1)),x_2)$,
$r_2: C(x_1,x_2)\gt{b}x_1$, $r_3: C(x_1,x_2)\gt{c}x_2$ we have that 
$r_1$, $r_1r_2$, and $r_1r_2r_3$ are stairs, 
since $A(x_1,x_2)\trans{r_1}C(C(x_2,B(x_2,x_1)),x_2)\trans{r_2}
C(x_2,B(x_2,x_1))\trans{r_3}B(x_2,x_1)$,
$r_1r_2r_2$ is no stair, since  $A(x_1,x_2)\trans{r_1r_2r_2}x_2$,
and $r_1r_2r_1$ is no stair since 
$A(x_1,x_2)\trans{r_1r_2}C(x_2,B(x_2,x_1))$ and $r_1$ is not enabled in 
$C(..,..)$.

We put $\varin(E)=\{x\in\var\mid x$ occurs in $E\}$.
A substitution $\sigma$ is called a \emph{stair substitution} if 
the sets
$\support(\sigma)$
and $\varin(x_i\sigma)$, $i\in[1,m]$, are subsets of 
$\{x_1,x_2,\dots,x_m\}$.
We note that 
each stair $w\in\calR^+$ determines a path 
$A(x_1,\dots,x_m)\gt{w}B(x_1,\dots,x_m)\sigma$ where
$\sigma$ is a stair substitution.
(E.g.,
$A(x_1,x_2)\trans{r_1r_2}C(x_2,B(x_2,x_1))$ can be written as
$A(x_1,x_2)\trans{r_1r_2}C(x_1,x_2)\sigma$ where
$\sigma=[x_1/x_2,x_2/B(x_2,x_1)]$.) 
In fact, the above defined stair substitutions are more general; 
e.g., the empty-support substitution is also a stair substitution.

For a stair substitution $\sigma$
			we define: 
\begin{itemize}			
			\item
$\surv(\sigma)=\{x_i\mid x_i$ occurs in $x_j\sigma$ for some
$j\in[1,m]\}$, and
\item
$\rstick(\sigma)=\{x_i\mid x_i=x_j\sigma$  for some
$j\in[1,m]\}$.
\end{itemize}			 		
Hence
$\rstick(\sigma)\subseteq\surv(\sigma)\subseteq\{x_1,\dots,x_m\}$.
We note that $x_i\in\surv(\sigma)$ iff $x_i$ ``survives'' applying $\sigma$ to
$A(x_1,\dots,x_m)$, i.e., $x_i$ occurs in $A(x_1,\dots,x_m)\sigma$; 
we have $x_i\in\rstick(\sigma)$ iff $x_i$ ``sticks to the root'', i.e.,
is a root-successor in
$A(x_1,\dots,x_m)\sigma$. 

A \emph{stair substitution} $\sigma$ is \emph{colour-idempotent} if 
for all $i,j\in[1,m]$ we have:

\begin{enumerate}
	\item
		$x_i\in\rstick(\sigma)$ entails that $x_i\sigma=x_i$, and 	
	\item
$x_i\in\surv(\sigma)$ entails that $x_i$ occurs in $x_j\sigma$ for some
$x_j\in\surv(\sigma)$.
\end{enumerate}

\emph{Remark.} The word ``colour'' anticipates a later
use of Ramsey's theorem. 
The word ``idempotent'' refers to the fact (shown below)
that the above conditions $1$ and $2$ are sufficient to guarantee 
that $\surv(\sigma\sigma)=\surv(\sigma)$
and $\rstick(\sigma\sigma)=\rstick(\sigma)$.
We could explicitly build a concrete finite semigroup of colours (of stair
substitutions) but this is not necessary for our proof. We
only remark that for technical reasons
the colour associated with $\sigma$ before
Proposition~\ref{prop:strongsubseq} is finer than
$(\surv(\sigma),\rstick(\sigma))$.

\medskip

\begin{figure}[t]
\centering
\includegraphics[scale=0.37]{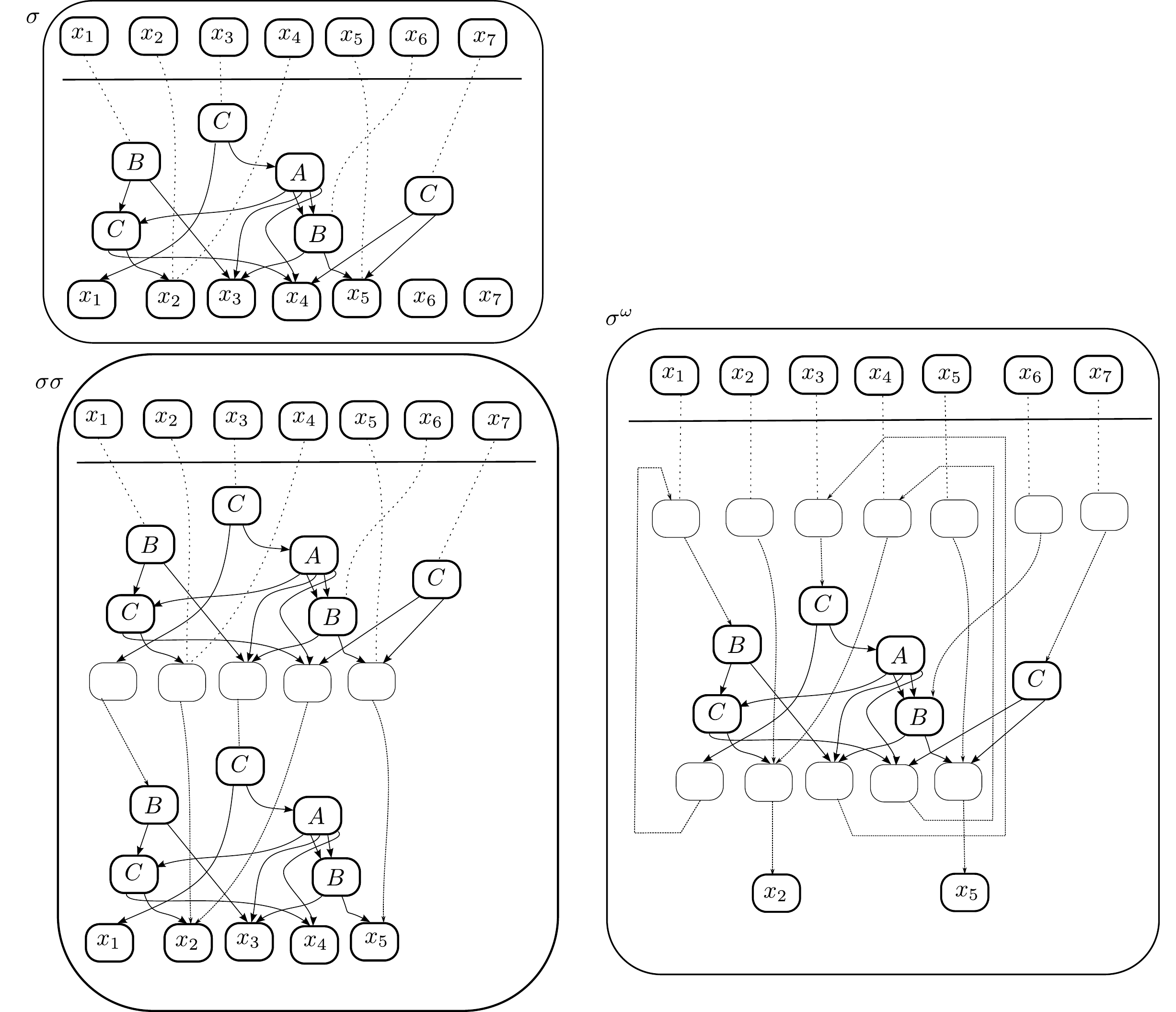}
\caption{A colour-idempotent stair substitution $\sigma$, 
	 and its limit $\sigma^\omega$}\label{fig:infsubstit}
\end{figure}
\emph{Example}.
Fig.~\ref{fig:infsubstit} (top-left) depicts a stair substitution
\[\sigma=[x_1/F_1, 
x_3/F_3,
x_4/F_4, x_6/F_6, x_7/F_7];\] we assume $m=7$ but we also use nonterminals with smaller
arities for simplicity.
We have 	$F_1=B(C(x_4,x_2),x_3)$, 
$F_3=C(x_1,A(C(x_4,x_2),B(x_3,x_5),x_3,B(x_3,x_5),x_4))$,
$F_4=x_2$,
$F_6=B(x_3,x_5)$,
$F_7=C(x_4,x_5)$.
It is also depicted explicitly 
that $x_2\sigma=x_2$ and $x_5\sigma=x_5$;
each dashed line connects a variable $x_i$ (above the bar) with the root of 
the term $x_i\sigma$. 

We can easily check that 
$\surv(\sigma)=\{x_1,x_2,x_3,x_4,x_5\}$ and 
$\rstick(\sigma)=\{x_2,x_5\}$, and that $\sigma$ is
colour-idempotent ($x_2\sigma=x_2$, $x_5\sigma=x_5$,
$x_1$ occurs in $x_3\sigma$, and both $x_3,x_4$ occur, e.g., 
in $x_1\sigma$).

Fig.~\ref{fig:infsubstit} also depicts the substitution
$\sigma\sigma$  (bottom-left) and the substitution $\sigma^\omega$ 
(right).
 Here we use an auxiliary device, namely 
some ``fictitious'' nodes that  are not
labelled with nonterminals or variables.
Such a node can be called 
a \emph{collector node}: it might ``collect'' several incoming arcs
that are in reality 
deemed to proceed to the target specified by the (precisely one)
outgoing arc of the collector node. 

Fig.~\ref{fig:infsubstit} can also serve for illustrating the next
proposition.

\begin{proposition}\label{prop:substidemp}
If $\sigma$ is a colour-idempotent stair substitution, then the
following conditions hold:
\begin{enumerate}
	\item
	$\surv(\sigma^k)=\surv(\sigma)$ 
and  $\rstick(\sigma^k)=\rstick(\sigma)$ for all $k\in\Natpos$;
\item
for each $x_i\in\surv(\sigma)\smallsetminus\rstick(\sigma)$, all
occurrences of $x_i$ in $x_j\sigma^k$, for $j\in[1,m]$ and
$k\in\Natpos$, have depths at least $k$ (if there are any such
occurrences).
\item
$\surv(\sigma^\omega)=\rstick(\sigma^\omega)=\rstick(\sigma)$.
\end{enumerate}		
\end{proposition}
\begin{proof}
1. By induction on $k$, the case $k=1$ being trivial.
We note that $x_i\in\surv(\sigma^{k+1})$ iff 
$x_i$ occurs  
in $x_j\sigma$ for some $x_j\in\surv(\sigma^k)$,
and $\surv(\sigma^k)=\surv(\sigma)$ by the induction hypothesis.
Since $\sigma$ is colour-idempotent, the condition 
``$x_i$ occurs  
in $x_j\sigma$ for some $x_j\in\surv(\sigma)$'' is equivalent
to ``$x_i\in\surv(\sigma)$''. Hence
$\surv(\sigma^{k+1})=\surv(\sigma)$.

Similarly, $x_i\in\rstick(\sigma^{k+1})$ iff $x_j\sigma=x_i$ for some 
$x_j\in\rstick(\sigma^k)=\rstick(\sigma)$. 
The condition ``$x_j\sigma=x_i$ for some 
$x_j\in\rstick(\sigma)$''  is equivalent
to ``$x_i\in\rstick(\sigma)$'' (since $\sigma$ is colour-idempotent).
 Hence
$\rstick(\sigma^{k+1})=\rstick(\sigma)$.

2. For $k=1$ the claim is trivial (by the definition of
$\rstick(\sigma)$).
The depth of any respective occurrence of $x_i$ in $x_j\sigma^{k+1}$
is the sum of the depth of an occurrence of $x_\ell$ in 
$x_j\sigma^{k}$ and the depth of $x_i$ in $x_\ell\sigma$ (for some
$\ell\in[1,m]$). The second depth is at least $1$ (since 
$x_\ell\sigma=x_i$ would entail
$x_i\in\rstick(\sigma)$); 
 the first
depth is at least $k$ by the induction hypothesis,
since
$x_\ell\in\surv(\sigma^k)=\surv(\sigma)$ and $x_\ell\not\in\rstick(\sigma)$
(otherwise $x_\ell\sigma=x_\ell$ due to the colour-idempotency of
$\sigma$). Hence the depth of the respective occurrence of $x_i$ in $x_j\sigma^{k+1}$
is at least $k{+}1$.

3. Due to the (colour-idempotency) condition 
``$x_i\in\rstick(\sigma)$ entails
$x_i\sigma=x_i$'' (i.e., $x_j\sigma=x_i$ entails
$x_i\sigma=x_i$), the
substitution $\sigma^\omega$ is clearly well-defined: 
if $x_j\sigma=x_i$, then $x_j\sigma^\omega=x_i$.
Hence each variable $x_i\in\rstick(\sigma)$ ``survives'' the
application of $\sigma^\omega$ to $A(x_1,\dots,x_m)$, having one
occurrence as the $i$th root-successor in 
$A(x_1,\dots,x_m)\sigma^\omega$.
No other variables ``survive'', as can be easily deduced 
from Points $1,2$. (Suppose $x_i$ occurs in $A(x_1,\dots,x_m)\sigma^\omega$ in
depth $k$, and write $A(x_1,\dots,x_m)\sigma^\omega$ as 
$A(x_1,\dots,x_m)\sigma^{k+1}\sigma^\omega$; hence $x_i$ occurs in 
$x_j\sigma^\omega$ for some $x_j$ occurring in 
$A(x_1,\dots,x_m)\sigma^{k+1}$ in depth at most $k$. Points $1,2$
imply that
$x_j\in\rstick(\sigma)$; hence $x_j\sigma^\omega=x_j$, and thus
$x_i=x_j\in\rstick(\sigma)$.)
\end{proof}

We say that a \emph{stair} $w\in\calR^+$ is \emph{colour-idempotent}
if
$A(x_{1},\dots,x_{m})\gt{w}A(x_1,\dots,x_m)\sigma$ for some 
$A\in\calN$ and some colour-idempotent stair substitution $\sigma$.
For such $w$ 
we have the infinite path
\begin{center}
$A(x_{1},\dots,x_{m})\gt{w}A(x_{1},\dots,x_{m})\sigma\gt{w}
A(x_{1},\dots,x_{m})\sigma^2\gt{w}A(x_{1},\dots,x_{m})\sigma^3\gt{w}
\cdots$.
\end{center}
The term $A(x_{1},\dots,x_{m})\sigma^\omega$
 can be
naturally seen as the respective ``limit''.

\subparagraph*{Candidates for witnesses of bisim-infiniteness.}

Given a grammar $\calG=(\calN,\act,\calR)$, 
	by a \emph{candidate for a witness of
	bisim-infiniteness of a term $E_0$}, or by a \emph{candidate
	for $E_0$} for
	short, we mean
	a pair $(u,w)$ where $u\in\calR^*$, $w\in\calR^+$,
	$E_0\gt{uw}$, and $w$ is a colour-idempotent stair.

\begin{figure}[ht]
\centering
\includegraphics[scale=0.4]{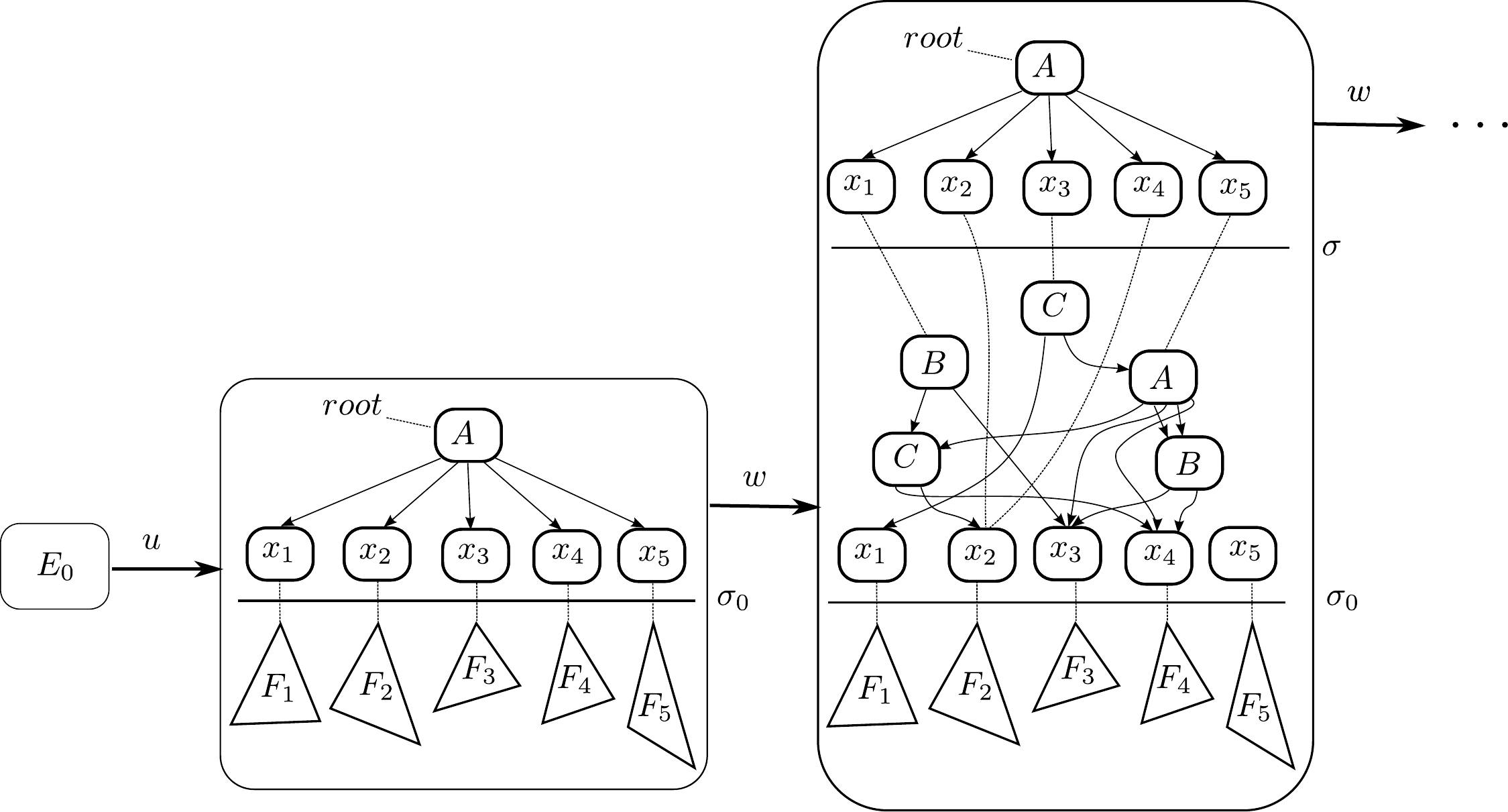}
\caption{Candidate $(u,w)$ induces the path
$E_0\gt{u}A(x_1,\dots,x_m)\sigma_0\gt{w}A(x_{1},\dots,x_{m})\sigma\sigma_0\gt{w}\cdots$}\label{fig:witnessfinal}
\end{figure}

\begin{figure}[h!]
\centering
\includegraphics[scale=0.4]{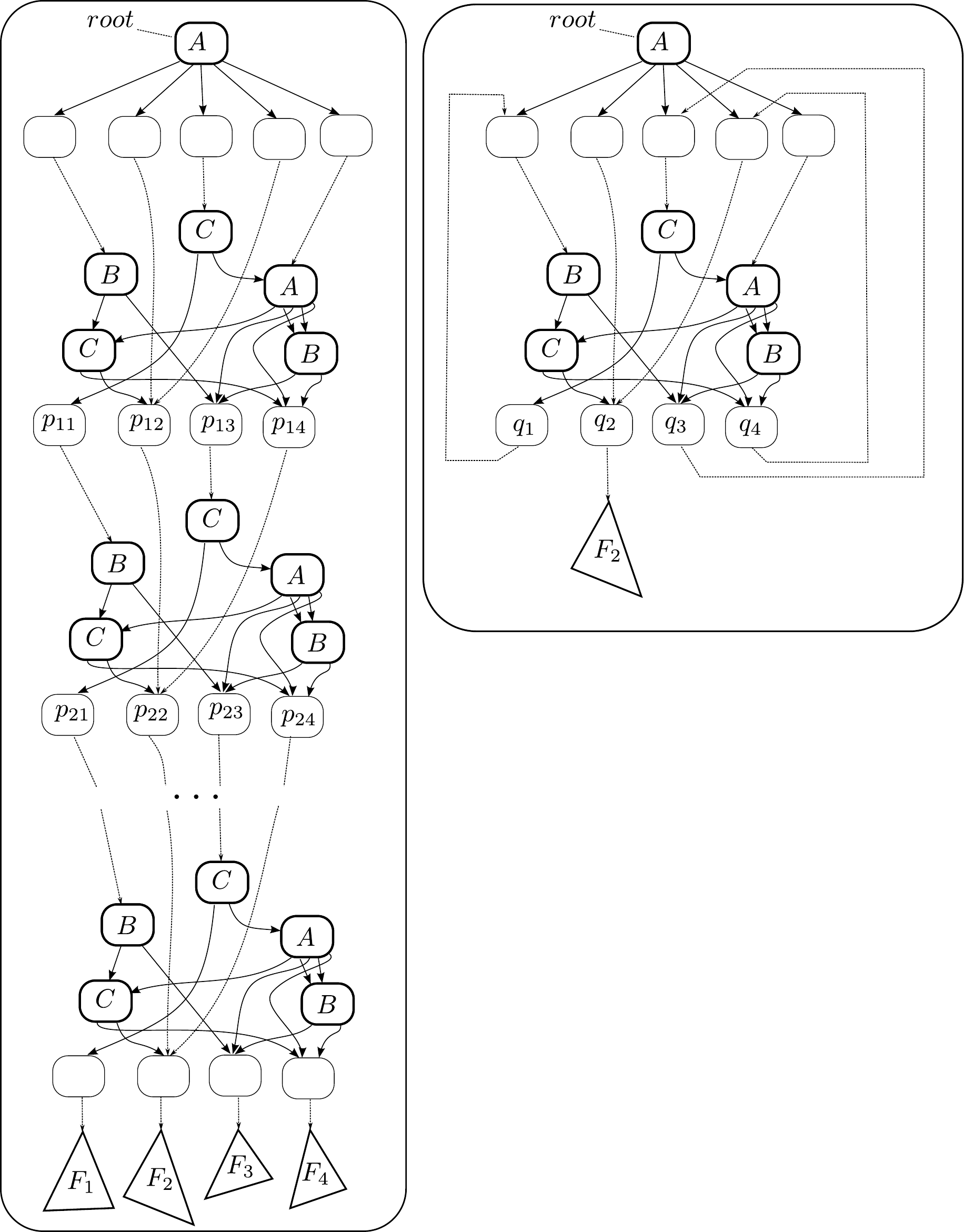}
\caption{$A(x_1,\dots,x_m)\sigma^j\sigma_0$ (left)
	and 
$A(x_1,\dots,x_m)\sigma^\omega\sigma_0$ (right)}\label{fig:witnesslimfinal}
\end{figure}

	For a candidate $(u,w)$ for $E_0$ we thus have
	$E_0\gt{u}A(x_1,\dots,x_m)\sigma_0$ and
$A(x_{1},\dots,x_{m})\gt{w}A(x_1,\dots,x_m)\sigma$
for some nonterminal $A$ and some substitutions $\sigma_0,\sigma$, 
where $\sigma$ is colour-idempotent; moreover,
there is the corresponding infinite path
$$E_0\gt{u}A(x_{1},\dots,x_{m})\sigma_0\gt{w}A(x_{1},\dots,x_{m})\sigma\sigma_0\gt{w}A(x_{1},\dots,x_{m})\sigma^2\sigma_0\gt{w}\cdots.$$
An example is depicted in Fig.~\ref{fig:witnessfinal}.
Fig.~\ref{fig:witnesslimfinal} then depicts 
$A(x_{1},\dots,x_{m})\sigma^j\sigma_0$ for some
$j\geq 3$ (left) and the limit $A(x_{1},\dots,x_{m})\sigma^\omega\sigma_0$
(right);
some of the auxiliary
collector nodes have special labels ($p_{ij}$, $q_i$) 
that we now ignore
(they serve for a later discussion).

We will now note in more detail how  
the terms $A(x_{1},\dots,x_{m})\sigma^j\sigma_0$
converge (syntactically and semantically) to the term
$A(x_{1},\dots,x_{m})\sigma^\omega\sigma_0$.

\subparagraph*{Tops of terms $A(x_{1},\dots,x_{m})\sigma^k\sigma_0$ converge to 
$A(x_{1},\dots,x_{m})\sigma^\omega\sigma_0$.} 
For a term $H$ and $d\in\Nat$, by $\topterm_d(H)$
(the \emph{$d$-top of $H$})
we refer to the tree
corresponding to 
$H$ up to depth $d$.
Hence 
$\topterm_0(H)$ is the tree consisting solely 
of the root labelled with $\termroot(H)$.
For $d>0$, we have
$\topterm_d(x_i)=\topterm_0(x_i)$, and
$\topterm_d(A(G_1,\dots,G_m))=A(\topterm_{d-1}(G_1),\dots,\topterm_{d-1}(G_m))$,
which denotes 
the (ordered labelled) tree
with the $A$-labelled root and with the (ordered) depth-$1$ subtrees
 $\topterm_{d-1}(G_1),\dots,\topterm_{d-1}(G_m)$.
(Hence $\topterm_d(H)$ is not a term in general, since it arises by 
``cutting-off''
the depth-$(d{+}1)$ subterm-occurrences.)
We also
define $\topterm_{-1}(H)$ as the ``empty tree'', and use the
consequence that  $\topterm_{-1}(H_1)=\topterm_{-1}(H_2)$ for all
$H_1,H_2$.

The next observation is trivial, due to the root-rewriting form
of 
 the transitions in the grammar-generated labelled
transition systems.

\begin{proposition}\label{prop:topequiv}
For any $k\in\Nat$,
	if $\topterm_{k-1}(H_1)=\topterm_{k-1}(H_2)$, then
	$H_1\sim_k H_2$.
\end{proposition}
\begin{proof}	
We proceed by induction on $k$; for $k=0$ the claim is trivial.
We assume $\topterm_k(H_1)=\topterm_k(H_2)$ for $k\geq 0$;
hence $\termroot(H_1)=\termroot(H_2)$ and thus the rules
	$r$ enabled in $H_1$ in the LTS  $\calL^{\ltsrul}_\calG$ are
	the same as the rules enabled in $H_2$.
If a transition	$H_1\gt{a}H'_1$ in  $\calL^{\ltsact}_\calG$
arises from
$H_1\gt{r}H'_1$ in $\calL^{\ltsrul}_\calG$, 
then $H_2\gt{r}H'_2$ in $\calL^{\ltsrul}_\calG$
gives rise to  $H_2\gt{a}H'_2$ in  $\calL^{\ltsact}_\calG$; we
obviously have
$\topterm_{k-1}(H'_1)=\topterm_{k-1}(H'_2)$, and thus
$H'_1\sim_k H'_2$ by the induction hypothesis.
Hence $\topterm_k(H_1)=\topterm_k(H_2)$ implies 
$H_1\sim_{k+1}H_2$.
\end{proof}

We now derive an easy consequence (for which
Fig.~\ref{fig:witnesslimfinal} can be
useful).

\begin{proposition}\label{prop:pumpreach}
Let $\sigma$ be a
colour-idempotent substitution, and $\sigma_0$ a substitution
(with $\support(\sigma_0)\subseteq\{x_1,\dots,x_m\}$).
For all $j\in[1,m]$ and $k\in\Natpos$ we have 
$x_j\sigma^k\sigma_0\sim_k x_j\sigma^\omega\sigma_0$.
Hence for all $A\in\calN$ and  $k\in\Natpos$ we have 
$A(x_1,\dots,x_m)\sigma^k\sigma_0\sim_{k+1} A(x_1,\dots,x_m)\sigma^\omega\sigma_0$.
\end{proposition}
\begin{proof}
Assuming $\sigma$ and $\sigma_0$, we fix 
 $j\in[1,m]$ and $k\in\Natpos$, and show that  
$\topterm_{k-1}(x_j\sigma^{k}\sigma_0)=
			\topterm_{k-1}(x_j\sigma^\omega\sigma_0)$.
We write $x_j\sigma^\omega\sigma_0$ as
$x_j\sigma^k\sigma^\omega\sigma_0$, and recall that the variables $x_i$
occurring in $x_j\sigma^k$ are from $\surv(\sigma^k)=\surv(\sigma)$
and all occurrences of
$x_i\in\surv(\sigma)\smallsetminus\rstick(\sigma)$ in $x_j\sigma^k$
are in depth $k$ at least (by Proposition~\ref{prop:substidemp}).
Moreover, for each $x_i\in\rstick(\sigma)$ we have
$x_i\sigma^\omega=x_i$ and thus
$x_i\sigma_0=x_i\sigma^\omega\sigma_0$. Hence we indeed have
$\topterm_{k-1}(x_j\sigma^{k}\sigma_0)=
			\topterm_{k-1}(x_j\sigma^\omega\sigma_0)$.
We thus also get that			
$\topterm_{k}(A(x_1,\dots,x_m)\sigma^{k}\sigma_0)=
			\topterm_{k}(A(x_1,\dots,x_m)\sigma^\omega\sigma_0)$.
			The claim thus follows by Proposition~\ref{prop:topequiv}.
\end{proof}

\subparagraph*{Checking if a candidate is a witness.}
A candidate $(u,w)$ for $E_0$, yielding the path
$E_0\gt{u}A(x_1,\dots,x_m)\sigma_0\gt{w}
A(x_1,\dots,x_m)\sigma\sigma_0\gt{w}A(x_1,\dots,x_m)\sigma^2\sigma_0\gt{w}
\cdots$,
and the term $\LIMIT=A(x_1,\dots,x_m)\sigma^\omega\sigma_0$,
is a 
\emph{witness} (of bisim-infiniteness) \emph{for $E_0$} if 
$A(x_1,\dots,x_m)\sigma^k\sigma_0\not\sim \LIMIT$
for infinitely many
$k\in\Nat$.
	
Since 
$\eqlevel(A(x_1,\dots,x_m)\sigma^k\sigma_0,\LIMIT)> k$
(as follows from Prop.~\ref{prop:pumpreach}), we then
have that for each $e\in\Nat$ there is $k\in\Nat$ such that
\begin{center}
$e<\eqlevel(A(x_1,\dots,x_m)\sigma^k\sigma_0,\LIMIT)<\omega$, 
\end{center}
and 
Prop.~\ref{prop:increaseqlevel} thus confirms that $E_0$ is
indeed bisim-infinite if it has a witness.

The existence of an algorithm  checking if
a candidate is a witness follows from
the next lemma (which we prove by using the fundamental fact captured 
by Theorem~\ref{thm:pdabisimdecid}, also using
the ``labelled collector nodes''
in Fig.~\ref{fig:witnesslimfinal} for illustration).

\begin{lemma}\label{lem:verifcand}
Given $A\in\calN$, a colour-idempotent substitution $\sigma$,
and a substitution $\overline{\sigma}_0$ with
$\support(\overline{\sigma}_0)\subseteq\rstick(\sigma)$,
	there is a
	computable number $e\in\Nat$ such that 
for the term $\LIMIT=A(x_1,\dots,x_m)\sigma^\omega\overline{\sigma}_0$
and any substitution $\sigma_0$ 
coinciding with $\overline{\sigma}_0$ on $\rstick(\sigma)$
	one of the following conditions holds:
	\begin{enumerate}	
 \item
$A(x_1,\dots,x_m)\sigma^k\sigma_0\not\sim\LIMIT$
 for all integers $k\geq e$, or
		\item
$A(x_1,\dots,x_m)\sigma^k\sigma_0\sim\LIMIT$
 for all integers $k\geq e$.
\end{enumerate}	 
\end{lemma}	
To verify that a candidate $(u,w)$, where
$E_0\gt{u}A(x_1,\dots,x_m)\sigma_0\gt{w}A(x_1,\dots,x_m)\sigma\sigma_0$,
is a witness for $E_0$, it thus suffices to
compute $e$ for $A$, $\sigma$, $\overline{\sigma}_0$ where
$\overline{\sigma}_0$ is the restriction of $\sigma_0$ to
$\rstick(\sigma)$, and show that
$A(x_1,\dots,x_m)\sigma^e\sigma_0\not\sim 
A(x_1,\dots,x_m)\sigma^\omega\overline{\sigma}_0$.

Now we prove the lemma.

\begin{proof}
We assume  $A\in\calN$, a colour-idempotent substitution $\sigma$,
and a substitution $\sigma_0$; we define 
$\overline{\sigma}_0$ as the restriction of $\sigma_0$ to
$\rstick(\sigma)$ and put
$\LIMIT=A(x_1,\dots,x_m)\sigma^\omega\overline{\sigma}_0$
(hence $\LIMIT=A(x_1,\dots,x_m)\sigma^\omega\sigma_0$ since only 
$x_i\in\rstick(\sigma)$ occur in $A(x_1,\dots,x_m)\sigma^\omega$).

We can surely compute a number $d\in\Nat$ 
such that each 
$x_i\in\surv(\sigma)$ belongs to both of the (reachability) regions
$\region(A(x_1,\dots,x_m)\sigma,d)$ and  
$\region(A(x_1,\dots,x_m)\sigma\sigma,d)$
(recall~(\ref{eq:defregion})). (In Fig.~\ref{fig:witnesslimfinal} it
means that the terms whose roots are determined by the collector nodes labelled $p_{11},\cdots,p_{24}$ are
reachable within $d$ moves from the term $A(x_1,\dots,x_m)\sigma^j\sigma_0$.)

We define the test set 
$\calT=\{x_j\sigma^\omega\overline{\sigma}_0\mid x_j\in\surv(\sigma)\}$,
and put
\begin{center}
	$\maxtel=\sup(\EqLevels(\LIMIT,d,\calT))$
 \end{center}
 (recalling~(\ref{eq:defTEL})).
The set $\region(\LIMIT,d)$ and its subset $\calT$ are finite,
and easily constructible. (In Fig.~\ref{fig:witnesslimfinal},
the roots of the terms in $\calT$ are determined by the
collector nodes labelled $q_{1},\cdots,q_{4}$.)
Hence the number $\maxtel$ is finite, i.e., $\maxtel\in\{-1\}\cup\Nat$, and 
computable 
 by Theorem~\ref{thm:pdabisimdecid}.
We now put
\begin{center}
$e=\maxtel+2$
\end{center}
and show that this $e$ 
(computed from $A$, $\sigma$, $\overline{\sigma}_0$)
satisfies the claim.

In fact, it suffices to show that for each $k\geq e$ we have:
\begin{center}
$A(x_1,\dots,x_m)\sigma^{k}\sigma_0\not\sim \LIMIT$
iff $A(x_1,\dots,x_m)\sigma^{k+1}\sigma_0\not\sim \LIMIT$.
\end{center}
We start with assuming $k\geq e$ and 
$A(x_1,\dots,x_m)\sigma^{k}\sigma_0\not\sim \LIMIT$.
Written in another form, we thus have 
$\big(A(x_1,\dots,x_m)\sigma\big)\sigma^{k-1}\sigma_0\not\sim 
\big(A(x_1,\dots,x_m)\sigma\big)\sigma^\omega\sigma_0$.

By compositionality (Prop.~\ref{prop:congruence})
we then have 
$x_{j}\sigma^{k-1}\sigma_0\not\sim x_{j}\sigma^{\omega}\sigma_0$
	for some
		$x_j$ occurring in
		$A(x_1,\dots,x_m)\sigma$, i.e., 
		for some $x_{j}\in\surv(\sigma)$; 
		in fact,
		$x_j\in\surv(\sigma)\smallsetminus\rstick(\sigma)$
		(since $x_i\sigma^{k-1}=x_i\sigma^\omega=x_i$ for
		$x_i\in\rstick(\sigma)$). 
		We fix such 
		$x_j$ and recall that it also occurs
		in $A(x_1,\dots,x_m)\sigma\sigma$ (since
		$\surv(\sigma\sigma)=\surv(\sigma)$).

Since
$\eqlevel(x_{j}\sigma^{k-1}\sigma_0,x_{j}\sigma^\omega\sigma_0)\geq
k{-}1$ (by Prop.~\ref{prop:pumpreach}) and $k\geq e$, we have 
\begin{equation}\label{eq:mezimaxtel}
\maxtel<\eqlevel(x_{j}\sigma^{k-1}\sigma_0,x_{j}\sigma^\omega\sigma_0)<\omega;
\end{equation}	
we recall that $x_{j}\sigma^\omega\sigma_0$ belongs to the test set $\calT$.
Our choice of $d$ guarantees that $x_j$ belongs to
		$\region(A(x_1,\dots,x_m)\sigma\sigma,d)$, and thus
		\begin{center}
		$x_{j}\sigma^{k-1}\sigma_0$ belongs to 
	$\region(A(x_1,\dots,x_m)\sigma^{k+1}\sigma_0,d)$.
		\end{center}		
This implies that 
$A(x_1,\dots,x_m)\sigma^{k+1}\sigma_0\not\sim\LIMIT$ (by
Prop.~\ref{prop:regiontest}).

Similarly, if $A(x_1,\dots,x_m)\sigma^{k+1}\sigma_0\not\sim\LIMIT$
(for $k\geq e$), i.e., 
$\big(A(x_1,\dots,x_m)\sigma\sigma\big)\sigma^{k-1}\sigma_0\not\sim
\big(A(x_1,\dots,x_m)\sigma\sigma\big)\sigma^{\omega}\sigma_0$,
there is $x_j\in\surv(\sigma\sigma)=\surv(\sigma)$ for
which~(\ref{eq:mezimaxtel}) holds.
Since $x_j$ is in 	$\region(A(x_1,\dots,x_m)\sigma,d)$,
the term $x_{j}\sigma^{k-1}\sigma_0$ belongs to 
	$\region(A(x_1,\dots,x_m)\sigma^{k}\sigma_0,d)$, and thus
$A(x_1,\dots,x_m)\sigma^{k}\sigma_0\not\sim\LIMIT$.
\end{proof}

\subsubsection{Each bisim-infinite term has a 
witness}\label{subsec:infhaswitness}
The last ingredient of the proof of Theorem~\ref{th:decidsemfinit}
is Lemma~\ref{lem:bisinfhaswitness} formulated and proven at the end of
this section.
The rough idea is that for each bisim-infinite term $E_0$ there is
an infinite path 
\[E_0\gt{u}H_0\gt{w_1}H_1\gt{w_2}H_2\gt{w_3}\cdots\]
where 
$H_i\not\sim H_j$ for all $i\neq j$, and 
$w_i$ are stairs (for all $i\in\Natpos$) that might be even chosen
from a finite set of ``simple'' stairs.
The infinite Ramsey theorem will easily yield that infinitely many sequences 
$w_iw_{i+1}\cdots w_j$ are colour-idempotent stairs (with the same
```colour'').
By a more
detailed analysis (and another use of Lemma~\ref{lem:verifcand})
we will
derive a contradiction 
when assuming that $E_0$ has no witness. 

We assume a fixed grammar $\calG=(\calN,\act,\calR)$ and first define a few
notions.

\subparagraph*{Canonical (witness) sequences.}
\begin{figure}[t]
\centering
\includegraphics[scale=0.32]{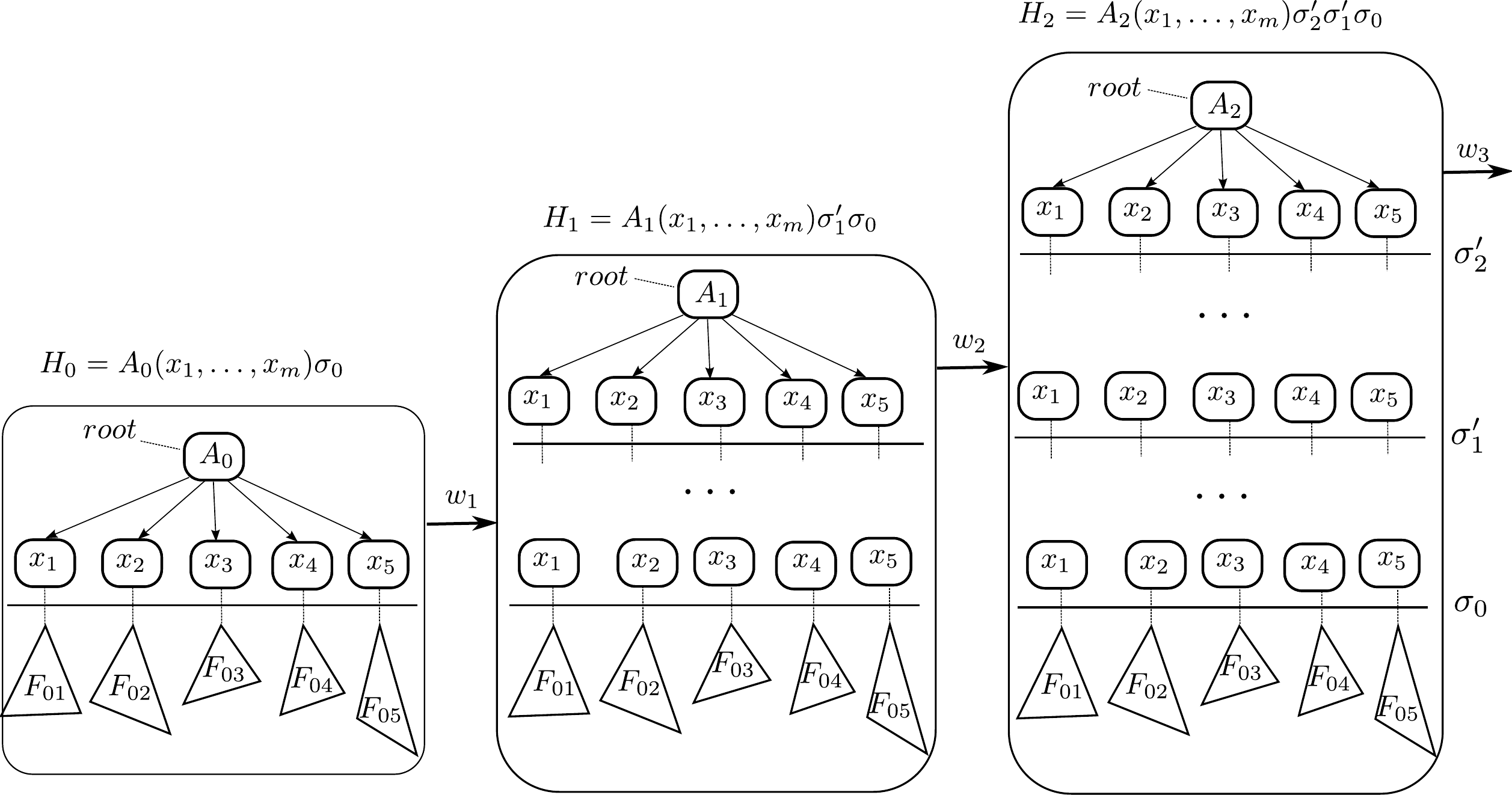}
\caption{Depiction of
$H_0\gt{w_1}H_1\gt{w_2}H_2\gt{w_3}\cdots$}\label{fig:canonseq}
\end{figure}
\begin{itemize}
		\item
For each rule $r: A(x_1,\dots,x_m)\gt{a}E$ in $\calR$ and each 
subterm $F$ of $E$ with
$\termroot(F)\in\calN$  (i.e., $F\not\in\var$)
we fix a shortest sequence 
$u_{(r,F)}=r_1r_2\cdots r_k\in\calR^+$ such that $r_1=r$ and
$A(x_1,\dots,x_m)\trans{r_1}E\trans{r_2\cdots r_k}F$.
Sequences $u_{(r,F)}$ are obviously stairs,
and we call them the \emph{canonical simple stairs}. 
\item
	A stair $w\in\calR^+$ is a \emph{canonical stair} if 
	$w=u_1u_2\cdots u_\ell$ where $\ell\in\Natpos$ and 
	$u_i$, $i\in[1,\ell]$, are canonical simple stairs.
\item
An infinite sequence $H_0$, $w_1$, $H_1$, $w_2$, $H_2$, $w_3$,
$\ldots$ (where $H_i\in\trees_\calN$ and $w_i\in\calR^+$) is a
\emph{canonical sequence} if $H_i\trans{w_{i+1}}H_{i+1}$ and $w_{i+1}$ is
a canonical stair, for each $i\in\Nat$.
(In Fig.~\ref{fig:canonseq} we can see a depiction of such a sequence.)
\item
A canonical sequence $H_0$, $w_1$, $H_1$, $w_2$, $H_2$, $w_3$,
$\ldots$ is a \emph{witness sequence} if $H_i\not\sim H_j$ for all
$i,j\in\Nat$ where $i\neq j$.
\item 
Given a canonical sequence $\seq=H_0, w_1, H_1, w_2, H_2, w_3,
\ldots$,  we say that a sequence 
$\seq'=H'_0$, $w'_1$, $H'_1$, $w'_2$, $H'_2$, $w'_3$,
$\ldots$
is a \emph{subsequence of} $\seq$ if there are $0\leq i_0<i_1<i_2<\cdots$ such
that $H'_j=H_{i_j}$
and $w'_{j+1}=w_{i_j+1}w_{i_j+2}\dots w_{i_{j+1}}$, 
for each $j\in\Nat$.
\item
	A \emph{canonical sequence} $\seq=H_0, w_1, H_1, w_2, H_2, w_3,
	\ldots$, is \emph{reachable from} a term $E_0$ if 
	$E_0\gt{u}H_0$ for some $u\in\calR^*$.
\end{itemize}			
\begin{proposition}\label{prop:easysubseq}
Let $\seq'$ be a subsequence of a canonical sequence $\seq$.
Then $\seq'$ is 
a canonical sequence; 
moreover, if $\seq$ is a witness sequence,
then $\seq'$ is a witness sequence, and if 
$\seq$ is reachable from $E_0$, then $\seq'$ is reachable from $E_0$.
\end{proposition}
\begin{proof}
It is obvious, once we note that if $H\gt{w}H'\gt{w'}H''$
	where $w$ and $w'$ are canonical stairs, then $ww'$ is a
	canonical stair.
\end{proof}	

\begin{proposition}\label{prop:witnessseq}
If a term $E_0$ is bisim-infinite, then there 
is a witness sequence $\seq$ that is reachable from $E_0$.
Moreover, there is such
$\seq=H_0, w_1, H_1, w_2, H_2, w_3,\ldots$ 
where $w_i$, $i\in\Natpos$, are canonical simple stairs.
\end{proposition}	
\begin{proof}
We assume a bisim-infinite term $E_0$ and  
 fix an infinite path
\begin{equation}\label{eq:firstinfpath}
E_0\gt{r_1}E_1\gt{r_2}E_2\gt{r_3}\cdots
\end{equation}
in the LTS $\calL^\ltsrul_\calG$ such that $E_i\not\sim E_j$ (in
$\calL^\ltsact_\calG$)
for all $i\neq j$; the existence of such a path follows from 
Prop.~\ref{prop:infinbisimchange}.

We show that there is the least
$i_0\in\Nat$
such that $r_{i_0+1}r_{i_0+2}\dots r_{i_0+\ell}$ is a stair for each
$\ell\in\Natpos$. If there was no such $i_0$, we would have 
an infinite sequence
$0=j_0<j_1<j_2<\cdots$ where $E_{j_{k+1}}$ is a depth-$1$ subterm of $E_{j_k}$
for each $k\in\Nat$ (since  $E_{j_k}\trans{r_{j_k+1}r_{j_k+2}\cdots
r_{j_{k+1}}}E_{j_{k+1}}$ sinks to a root-successor in $E_{j_k}$);
since
$E_0$ is a regular term,
it has only finitely many subterms, 
and we would thus have $E_i=E_j$ (hence $E_i\sim E_j$) for some $i\neq j$.

Having defined $i_0, i_1,\dots, i_j$ (for some
$j\geq 0$),
we define $i_{j+1}$ as the least
number $i$ 
  such
that $i_{j}<i$
and $r_{i+1}r_{i+2}\dots r_{i+\ell}$ is a stair for each
	$\ell\in\Natpos$. There must be such $i$, since otherwise 
	$E_{i_j}\trans{r_{i_j+1}r_{i_j+2}\cdots
	r_{i_j+\ell}}$ would sink to a root-successor in 
$E_{i_j}$ for some $\ell\in\Natpos$, which contradicts with the choice of
$i_j$.

For each $j\in\Nat$ we put $H_j=E_{i_j}$ and 
$w_{j+1}=r_{i_j+1}r_{i_j+2}\cdots r_{i_{j+1}}$; hence
the (infinite) suffix of the path~(\ref{eq:firstinfpath}) 
that 
starts with $E_{i_0}$ can be presented as
\[H_0\gt{w_1}H_1\gt{w_2}H_2\gt{w_3}\cdots\,.\]
By the choice of~(\ref{eq:firstinfpath}) 
we have  $H_i\not\sim H_j$ for
 $i\neq j$. 
By the definition of $i_0, i_1, i_2,\dots$, all $w_j\in\calR^+$ are stairs, but they might not be
canonical (simple) stairs. 
For each $j\in\Nat$ we can write $H_j\trans{w_{j+1}}H_{j+1}$
in the form $A(x_1,\dots,x_m)\sigma\trans{w_{j+1}}
F\sigma$ where $A=\termroot(H_j)$ and
$A(x_1,\dots,x_m)\trans{w_{j+1}}F$; in more detail, we have
$A(x_1,\dots,x_m)\trans{r_{i_j+1}}E\trans{r_{i_j+2}\cdots r_{i_{j+1}}}F$
where the rule $r_{i_j+1}\in \calR$ is of the form 
$A(x_1,\dots,x_m)\gt{a}E$. Moreover, by the definition of $i_j$ and $i_{j+1}$ 
we must have that $F$ is
a subterm of $E$ and  $F\not\in\var$ (i.e., $\termroot(F)\in\calN$).
Hence we have $H_j\trans{w'_{j+1}}H_{j+1}$ for the respective
canonical simple stair $w'_{j+1}=u_{(r,F)}$ where $r=r_{i_j+1}$.
By replacing all $w_{j+1}$ with the respective canonical simple stairs
$w'_{j+1}$ we get the desired sequence $\seq$.
\end{proof}

\subparagraph*{%Colours of substitutions and stairs, 
(Strong) monochromatic canonical sequences.}
We now build on the notions
$\rstick(\sigma)\subseteq\surv(\sigma)\subseteq\{x_1,\dots,x_m\}$  that
we introduced for stair substitutions $\sigma$ earlier.
(We recall that $\surv(\sigma)$ consists of the variables occurring in
$A(x_1,\dots,x_m)\sigma$, while $x_i\in\rstick(\sigma)$ are even
root-successors in $A(x_1,\dots,x_m)\sigma$.)

\begin{itemize}
	\item
		For a stair substitution $\sigma$ we define its colour as
\begin{center}
	$\colsub(\sigma)=\big(\surv(\sigma),(\rs_1,\rs_2,\dots,\rs_m)\big)$ 
\end{center}
where $\rs_i=\{x_j\mid x_j\sigma=x_i\}$, for $i\in[1,m]$.
Hence $\rstick(\sigma)=\{x_i\mid \rs_i\neq \emptyset\}$.
	\item
For a stair $w\in\calR^+$
where $A(x_1,\dots,x_m)\gt{w}B(x_1,\dots,x_m)\sigma$
	(with $\support(\sigma)\subseteq\{x_1,x_2,\dots,x_m\}$)
we define its colour as
\begin{center}
	$\colstair(w)=\big(A,B,\colsub(\sigma)\big)$ .
\end{center}
\item
If $\seq=H_0,w_1,H_1,w_2,H_2,w_3,\dots$ is a canonical sequence, then for
$i<j$ ($i,j\in\Nat$) we put $\colpair_\seq(i,j)=\colstair(w_{i+1}w_{i+2}\cdots
w_{j})$.
\item
A canonical sequence $\seq$
is \emph{a monochromatic sequence} if there is a ``colour''
$\textsc{c}=\big(A,A,(\sv,(\rs_1,\dots,\rs_m)) \big)$
such that 
$\colpair_\seq(i,j)=\textsc{c}$ for all $i<j$.
(This could not hold for $\textsc{c}=\big(A,B,\dots\big)$ where $A\neq
B$.) In this case we put $\surv_\seq=\sv$ and $\rstick_\seq=\{x_i\mid
\rs_i\neq \emptyset\}$.
\item 
If $\seq$ is a monochromatic sequence $H_0,w_1,H_1,w_2,H_2,w_3,\dots$,
presented as
\[A(x_1,\dots,x_m)\sigma_0,w_1,A(x_1,\dots,x_m)\sigma_1,w_2,A(x_1,\dots,x_m)\sigma_2,w_3,\dots,\]
and we have $x_i\sigma_0\sim x_i\sigma_1\sim
x_i\sigma_2\sim\cdots$ for each $x_i\in\surv_\seq$, then $\seq$ is 
a \emph{strong monochromatic sequence}.
(For each $x_i\in \surv_\seq$ there is a fixed bisim-class such
that the $i$th
root-successor in $H_j$ is from this fixed class,  for each
$j\in\Nat$.)
\end{itemize}

\begin{proposition}\label{prop:strongsubseq}
If a bisim-infinite term $E_0$ has no witness
and $\seq$ is a canonical sequence 
reachable from $E_0$, then $\seq$ has a strong monochromatic subsequence.
\end{proposition}

\begin{proof}
We fix a bisim-infinite term $E_0$ that has no witness
(assuming such $E_0$ exists); we further fix 
 a canonical sequence $\seq$ reachable from $E_0$.
By the infinite Ramsey theorem there is a monochromatic subsequence
$\seq'$ of $\seq$. We will thus immediately assume that
$\seq=H_0,w_1,H_1,w_2,H_2,w_3,\dots$ is monochromatic,
that $E_0\gt{u}H_0$,
and that
\begin{center}
	$\colpair_\seq(i,j)=\textsc{c}=\big(A,A,(\surv_\seq,(\rs_1,\dots,\rs_m))\big)$
for all $i<j$ ($i,j\in\Nat$).
\end{center}
Hence $\colstair(w_{j}w_{j+1}\cdots w_{j+\ell})=\textsc{c}$ for all
$j\in\Natpos$ and $\ell\in\Nat$.

Let us write $H_j=A(x_1,\dots,x_m)\sigma_j$ for all $j\in\Nat$. In more
detail, $\sigma_{j+1}=\sigma'_{j+1}\sigma_j$ where 
$A(x_1,\dots,x_m)\trans{w_{j+1}}A(x_1,\dots,x_m)\sigma'_{j+1}$,
for all $j\in\Nat$;
hence $\sigma_j=\sigma'_j\sigma'_{j-1}\cdots\sigma'_{1}\sigma_0$. We
thus also have
\begin{center}
$\colsub(\sigma'_\ell\sigma'_{\ell-1}\cdots\sigma'_{j})=\big(\surv_\seq,(\rs_1,\dots,\rs_m)\big)$
for all $\ell\geq j\geq 1$,
\end{center}
which also entails that
$\rstick(\sigma'_\ell\sigma'_{\ell-1}\cdots\sigma'_{j})=\rstick_\seq=\{x_i\mid\rs_i\neq\emptyset\}$.

We now verify that $\sigma'_1$ is colour-idempotent; 
the definition requires two conditions:
\begin{enumerate}
\item
($x_i\in\rstick(\sigma'_1)$ entails that $x_i\sigma'_1=x_i$)

	Suppose $x_i\in \rstick(\sigma'_1)=\rstick_\seq$, hence 
$x_j\sigma'_1=x_i$ for some $j\in[1,m]$, and thus
$x_j\in\rs_i$.
Since
$\colsub(\sigma'_1)=\colsub(\sigma'_{2})=\colsub(\sigma'_{2}\sigma'_1)$,
 we also have $x_j\sigma'_{2}=x_i$ and 
 $x_j\sigma'_{2}\sigma'_{1}=x_i$, which entails 
 $x_i\sigma'_1=x_i$.
\item
	($x_i\in\surv(\sigma'_1)$ entails that $x_i$ occurs in
	$x_j\sigma'_1$ for some
	$x_j\in\surv(\sigma'_1)$)
	
	For $x_i\in\surv(\sigma'_1)$ we also have
 $x_i\in\surv(\sigma'_{2}\sigma'_1)$
 (since $\surv(\sigma'_1)=\surv(\sigma'_{2}\sigma'_1)=\surv_\seq$), hence there is
 $x_j\in\surv(\sigma'_{2})$ such that $x_i$ occurs in
 $x_j\sigma'_1$; since
 $\surv(\sigma'_{2})=\surv(\sigma'_1)$, 
 we have that $x_i$ occurs in $x_j\sigma'_1$ for some
 $x_j\in\surv(\sigma'_1)$.
 \end{enumerate}
The colour-idempotency claim on $\sigma'_1$ can be obviously generalized, but it
suffices for us to note that each nonempty $\rs_i$ (in the colour
$\textsc{c}$) contains $x_i$,
hence $x_i\sigma'_j=x_i$ for all $x_i\in\rstick_\seq$ and
$j\in\Natpos$.

Let $w=w_1$, $\sigma=\sigma'_1$, and 
$j\in\Nat$. We have shown that $w$ is a colour-idempotent stair,
hence the pair $(uw_1w_2\cdots w_j,w)$ is
a candidate for a witness of
bisim-infiniteness of $E_0$.
We consider the path
\begin{center}
$E_0\trans{uw_1w_2\cdots w_j}A(x_1,\dots,x_m)\sigma_j\gt{w}
A(x_1,\dots,x_m)\sigma\sigma_j\gt{w}
A(x_1,\dots,x_m)\sigma\sigma\sigma_j\gt{w}\cdots$
\end{center}
and the corresponding limit
$\LIMIT_j=A(x_1,\dots,x_m)\sigma^\omega\sigma_j$.

The set of variables occurring in the term
$A(x_1,\dots,x_m)\sigma^\omega$ 
is $\rstick(\sigma)=\rstick_\seq$
(recall Proposition~\ref{prop:substidemp}).
Since $x_i\sigma'_j\sigma'_{j-1}\cdots \sigma'_1=x_i$ for each
$x_i\in\rstick_\seq$,
we have 
\begin{center}
$\LIMIT_j=A(x_1,\dots,x_m)\sigma^\omega\sigma_j=
A(x_1,\dots,x_m)\sigma^\omega\sigma'_j\sigma'_{j-1}\cdots\sigma'_1\sigma_0=
A(x_1,\dots,x_m)\sigma^\omega\sigma_0$,
\end{center}
or $\LIMIT_j=A(x_1,\dots,x_m)\sigma^\omega\overline{\sigma}_0$ where
$\overline{\sigma}_0$ is the restriction of $\sigma_0$ to $\rstick_\seq$.
Hence 
$\LIMIT_0=\LIMIT_1=\LIMIT_2=\cdots$, and we thus 
write just $\LIMIT$ instead of $\LIMIT_j$.

Let $e\in\Nat$ be the value related to $A$, $\sigma$, and 
$\overline{\sigma}_0$
as in Lemma~\ref{lem:verifcand}.
Since we assume that $E_0$ has no witness, we have
$A(x_1,\dots,x_m)\sigma^e\sigma_j\sim\LIMIT$; this holds for all
$j\in\Nat$.

Each $x_i\in\surv_\seq$ occurs in 
$A(x_1,\dots,x_m)\sigma^e$
(by Proposition~\ref{prop:substidemp}, since $\sigma$ is
colour-idempotent), and there 
 is thus a (``sinking'') path
$A(x_1,\dots,x_m)\sigma^e\gt{v}x_i$.
Hence there is a bound, independent of
$j$, such that all terms $x_i\sigma_j$ where $x_i\in\surv_\seq$ (and
$j\in\Nat$), are in a bounded
(reachability) distance from 
$A(x_1,\dots,x_m)\sigma^e\sigma_j$.
Since $A(x_1,\dots,x_m)\sigma^e\sigma_j\sim\LIMIT$, every path 
$A(x_1,\dots,x_m)\sigma^e\sigma_j\gt{v}x_i\sigma_j$ must be matched by
a path $\LIMIT\gt{v'}G$ where $|v'|=|v|$ and $x_i\sigma_j\sim G$.
Hence the equivalence classes $[x_i\sigma_j]_\sim$, for
$x_i\in\surv_\seq$ and $j\in\Nat$, are
in a bounded distance from 
the class $[\LIMIT]_\sim$ (in the quotient LTS related to
$\calL^{\ltsact}_\calG$); due to finite
branching, the respective set $\calF$ of classes in this bounded
distance from $[\LIMIT]_\sim$ is finite.
The pigeonhole principle thus yields that $\seq$ indeed has a strong
monochromatic subsequence.
\end{proof}

\begin{lemma}\label{lem:bisinfhaswitness}
For each grammar $\calG$ and each bisim-infinite term $E_0$ there
is a 
witness (satisfying the condition $1$,
namely $A(x_1,\dots,x_m)\sigma^e\sigma_0\not\sim\LIMIT$, in Lemma~\ref{lem:verifcand}). 
\end{lemma}
\begin{proof}
Given a grammar $\calG=(\calN,\act,\calR)$, let $E_0$ be a bisim-infinite term; for the sake of contradiction we
assume that $E_0$ has no witness. 
We fix a witness sequence 
$$\seq=H_0,w_1,H_1,w_2,H_2,w_3,\ldots$$
where $H_0$ is reachable from $E_0$;
it exists by
Proposition~\ref{prop:witnessseq}.
We recall that $H_j\not\sim H_{j'}$ for $j\neq j'$ (since $\seq$ is a
witness sequence), and $w_{j}$, for each $j\in\Natpos$,
is a nonempty finite sequence of
canonical simple stairs for which $H_{j-1}\trans{w_{j}}H_{j}$;
this entails $H_0\trans{w_1}\cdots\trans{w_j}H_j$.

For each $j\in\Natpos$ we fix 
a shortest canonical stair $w'_{j}$ for which there is $H'_j$ such that 
$H_0\trans{w'_j}H'_j$ and $H'_j\sim H_j$.
All $w'_{j}$ ($j\in\Natpos$) are pairwise different finite sequences 
of canonical simple
stairs. (We have $w'_j\neq w'_{j'}$ for $j\neq j'$ since $H'_j\not\sim
H'_{j'}$.) 
By K\"onig's lemma we can thus easily derive that we can fix
 an infinite sequence $u_1u_2u_3\cdots$ of 
canonical simple stairs such that
each sequence  
$u_1u_2\cdots u_\ell$ (for $\ell\in\Nat$)
 is a prefix of $w'_j$ for infinitely many
 $j\in\Natpos$.

We now consider the canonical sequence $H_0, u_1, G_1, u_2, G_2,  \dots$,
which is reachable from $E_0$ (since $H_0$ is reachable from $E_0$);
we also write $H_0$ as $A(x_1,\dots,x_m)\sigma_0$.
Since $E_0$ has no witness, by Proposition~\ref{prop:strongsubseq} this sequence has a
strong monochromatic subsequence
\begin{center}
$\seq'=G_{i_0},v_1,G_{i_1},v_2,G_{i_2},v_3,\dots$,
\end{center}
where $v_{j+1}=u_{i_j+1}u_{i_j+2}\cdots u_{i_{j+1}}$, for each
$j\in\Nat$; we also put $v_0=u_1u_2\cdots u_{i_0}$.
(We do not exclude that $G_{i_j}\sim G_{i_{j'}}$ for $j\neq j'$.)
Hence there are $B\in\calN$ and stair substitutions
$\overline{\sigma}_0$,
 $\overline{\sigma}_1$, $\overline{\sigma}_2$, $\dots$ such that 
 $A(x_1,\dots,x_m)\trans{v_0}B(x_1,\dots,x_m)\overline{\sigma}_0$ and
 $B(x_1,\dots,x_m)\trans{v_{j+1}}B(x_1,\dots,x_m)\overline{\sigma}_{j+1}$
 (for all $j\in\Nat$);
the infinite path
$H_0\trans{v_0}G_{i_0}\trans{v_1}G_{i_1}\trans{v_2}\cdots$ can be thus
 written as
 \begin{equation}\label{eq:secondmonoseq}
A(x_1,\dots,x_m)\sigma_0\trans{v_0}B(x_1,\dots,x_m)\overline{\sigma}_0\sigma_0
	 \trans{v_1}B(x_1,\dots,x_m)\overline{\sigma}_1\overline{\sigma}_0\sigma_0
	 \trans{v_2}\cdots.
 \end{equation}
 We have $\surv(\overline{\sigma}_j)=\surv_{\seq'}$ for all
 $j\in\Natpos$, and for each $x\in\surv_{\seq'}$
 the bisim-class of the term
 $x\overline{\sigma}_j\overline{\sigma}_{j-1}\cdots\overline{\sigma}_0\sigma_0$
 ($j\in\Nat$) is independent of $j$ (since $\seq'$ is strong
 monochromatic).

By our choice of the sequence $u_1u_2u_3\dots$, we can fix some 
$j$ such that $w'_{j}=v_0v_1v_2 w$ for a (canonical)
stair $w$. Hence $H_0\trans{w'_{j}}H'_{j}$ 
can be written
as
 \begin{center}
	 $A(x_1,\dots,x_m)\sigma_0\trans{v_0}B(x_1,\dots,x_m)\overline{\sigma}_0\sigma_0
	 \trans{v_1v_2}B(x_1,\dots,x_m)\overline{\sigma}_2\overline{\sigma}_1\overline{\sigma}_0\sigma_0
	 \trans{w}H'_{j}$
 \end{center}
 where $B(x_1,\dots,x_m)\trans{w}C(x_1,\dots,x_m)\sigma$ and
$H'_{j}=
 C(x_1,\dots,x_m)\sigma
 \overline{\sigma}_2\overline{\sigma}_1\overline{\sigma}_0\sigma_0.$

We finish the proof by showing that $w'=v_0v_2 w$
(arising from $w'_{j}$ by omitting $v_1$) 
contradicts the length-minimality condition put on $w'_{j}$.
We obviously have $H_0\trans{w'}H'$ where 
$H'=C(x_1,\dots,x_m)\sigma\overline{\sigma}_2\overline{\sigma}_{0}\sigma_0$,
and it is thus sufficient to show 
that $H'_j\sim H'$, i.e.,
\begin{equation}\label{eq:shortereq}
C(x_1,\dots,x_m)\sigma
 \overline{\sigma}_2\overline{\sigma}_1\overline{\sigma}_0\sigma_0\sim
C(x_1,\dots,x_m)\sigma\overline{\sigma}_2\overline{\sigma}_{0}\sigma_0.
\end{equation}
 Since 
 $\varin\left(C(x_1,\dots,x_m)\sigma\right)\subseteq\{x_1,\dots,x_m\}$
and
$\surv(\overline{\sigma}_{2})=\bigcup_{i\in[1,m]}\varin(x_i\overline{\sigma}_{2})=\surv_{\seq'}$, we derive that
$\varin\left(C(x_1,\dots,x_m)\sigma\overline{\sigma}_2\right)\subseteq \surv_{\seq'}$.
For each $x\in
\varin\left(C(x_1,\dots,x_m)\sigma\overline{\sigma}_2\right)$ we thus
have $x\overline{\sigma}_{1}\overline{\sigma}_{0}\sigma_0\sim
x\overline{\sigma}_{0}\sigma_0$
(recall reasoning around~(\ref{eq:secondmonoseq}));
hence the claim~(\ref{eq:shortereq}) follows due to compositionality captured 
by Proposition~\ref{prop:congruence}. 
\end{proof}

\section{Additional Remarks}\label{sec:AddRem}

The idea of decision procedures for bisim-finiteness (or language regularity)
in the deterministic case studied 
in~\cite{DBLP:journals/iandc/Stearns67,DBLP:journals/jacm/Valiant75}
could be
roughly explained  in our framework 
as follows. For a deterministic grammar 
(with at most one rule $A(x_1,\dots,x_m)\gt{a}..$ for
each nonterminal $A$ and each action $a$),
if we have
\begin{center}
$E_0\gt{u}A(x_1,\dots,x_m)\sigma_0\gt{w}A(x_1,\dots,x_m)\sigma\sigma_0$
\end{center}
where $w$ is a colour-idempotent stair,
then either $A(x_1,\dots,x_m)\sigma_0\sim
A(x_1,\dots,x_m)\sigma\sigma_0$
(in which case $A(x_1,\dots,x_m)\sigma\sigma_0$ can be always safely replaced
with the smaller $A(x_1,\dots,x_m)\sigma_0$), or 
$A(x_1,\dots,x_m)\sigma_0\not\sim
A(x_1,\dots,x_m)\sigma\sigma_0$ and 
$E_0$ is  bisim-infinite.
This allows to derive a bound on the
size of the potential equivalent finite system,
and thus the decidability of the full equivalence
(Theorem~\ref{thm:pdabisimdecid})
is not needed here. 

In the case equivalent to \emph{normed} pushdown processes, the
regularity problem essentially coincides with the boundedness problem,
and is thus much simpler. (See, e.g.,~\cite{Srba:Roadmap:04} for
a further discussion.)

\section*{Appendix}

At the ends of Sections~\ref{sec:prelim} and~\ref{subsec:witnessfacts}
 the issues of transforming pushdown automata to
first-order grammars, of normalizing the grammars, and of unifying the
nonterminal arities
were mentioned.
We now deal with these issues in more detail.

\subsection*{Transforming pushdown automata to first-order grammars}

A \emph{pushdown automaton} (\emph{PDA})
is a tuple 
$M=(Q,\act,\Gamma,\Delta)$ of finite sets where the elements of 
$Q,\act,\Gamma$ are called 
\emph{control states}, 
\emph{actions} (or \emph{terminal letters}), and
\emph{stack symbols}, respectively; $\Delta$ contains \emph{transition
rules} of the form  $pY \gt{a}q\alpha$ where $p,q\in Q$, $Y\in\Gamma$,
$a\in \act\cup\{\varepsilon\}$, and $\alpha\in \Gamma^*$.
(We assume $\varepsilon\not\in\Sigma$.)
A PDA $M=(Q,\act,\Gamma,\Delta)$ generates the labelled transition system
\begin{center}
$\calL_M=(Q\times \Gamma^*,\act\cup\{\varepsilon\}, 
(\gt{a})_{a\in\act\cup\{\varepsilon\}})$
\end{center}
where 
each rule $pY \gt{a}q\alpha$ induces transitions 
$pY\beta \gt{a}q\alpha\beta$ for all $\beta\in \Gamma^*$.

\begin{figure}[ht]
\centering
\includegraphics[scale=0.43]{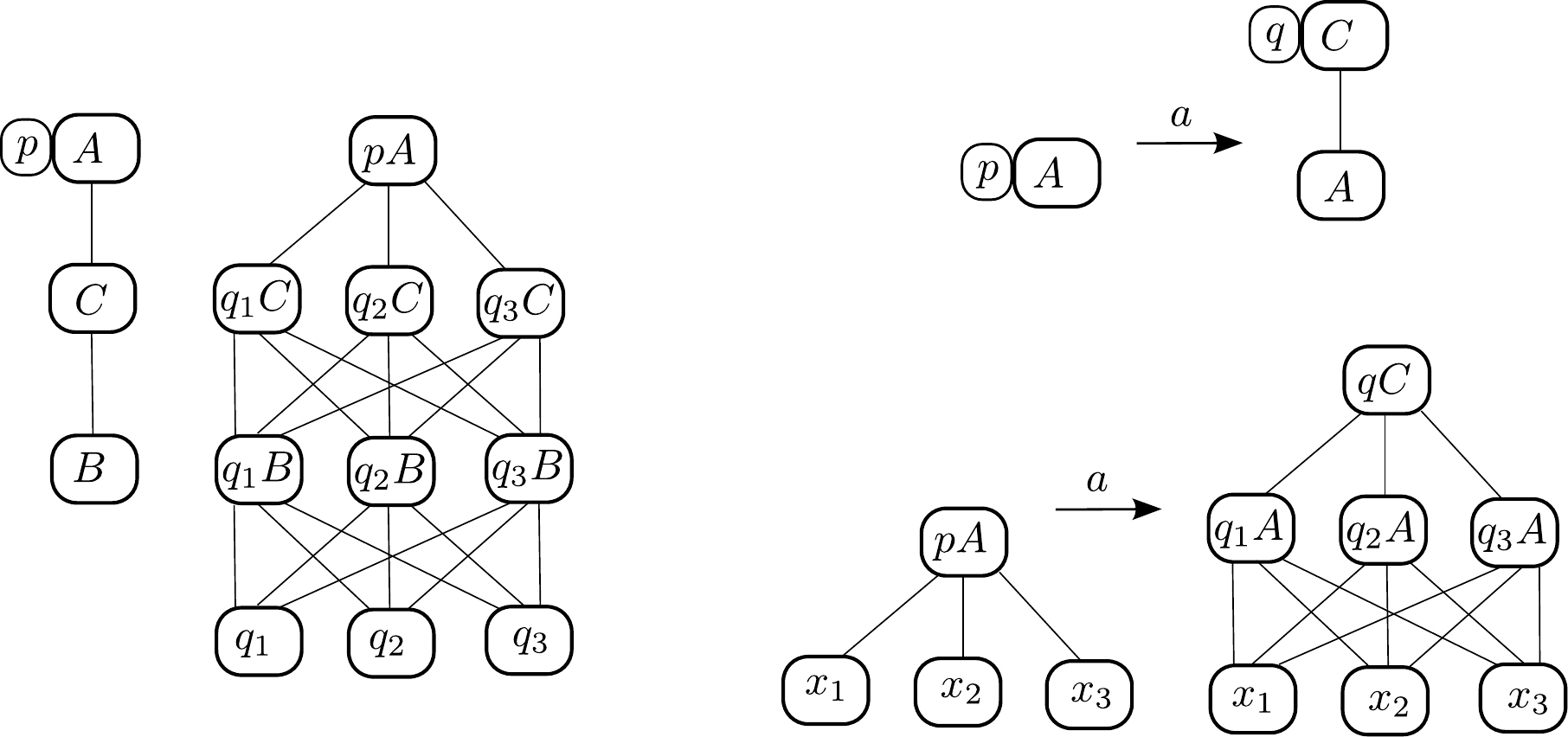}
	\caption{PDA configuration as a term (left), and transforming a
	rule (right)}\label{fig:pdatofo}
\end{figure}

Fig.~\ref{fig:pdatofo} (left) presents a PDA-configuration 
({i.e.}~a state in
$\calL_M$) $pACB$
as a term; here we assume that
$Q=\{q_1,q_2,q_3\}$. (The string  $pACB$, depicted on the left in a
convenient vertical form, is transformed into a term presented
by an acyclic graph in the figure.)
On the right in Fig.~\ref{fig:pdatofo} we can see a
transformation of a PDA-rule $pA\gt{a}qCA$ into a grammar-rule. 

Formally, for a PDA $M=(Q,\act,\Gamma,\Delta)$,
where $Q=\{q_1,q_2,\dots,q_m\}$, 
we can define the first-order grammar
$\calG_M=(\calN,\Sigma\cup\{\varepsilon\},\calR)$ where 
$\calN=Q\cup (Q\times \Gamma)$, with $\arity(q)=0$ and 
$\arity((q,X))=m$; the set $\calR$ is defined below.
We write $[q]$ and $[qY]$ for nonterminals $q$ and $(q,Y)$,
respectively, and we map 
each configuration $p\alpha$ to the term 
$\calT(p\alpha)$ by structural induction:
$\calT(p\varepsilon)=[p]$, and
$\calT(pY\alpha)=
[pY](\calT(q_1\alpha),\calT(q_2\alpha),\dots,\calT(q_m\alpha))$.

For a smooth transformation of rules we introduce a special
``stack variable'' $x$, and we put $\calT(q_ix)=x_i$ (for all $i\in[1,m]$).
A PDA-rule $pY\gt{a}q\alpha$ in $\Delta$ is transformed to the grammar
rule $\calT(pYx)\gt{a}\calT(q\alpha x)$ in $\calR$.
(Hence $pY\gt{a}q_i$ is transformed to $[pY](x_1,\dots,x_m)\gt{a}x_i$,
and  $pY\gt{a}qZ\alpha$ is transformed to 
$[pY](x_1,\dots,x_m)\gt{a}
[qZ](\calT(q_1\alpha x),\dots,\calT(q_m\alpha x)$.)

It is obvious that the LTS $\calL_M$ is isomorphic with the
restriction of the LTS $\calL^{\ltsact}_{\calG_M}$ to the states $\calT(p\alpha)$
where $p\alpha$ are configurations of $M$;
moreover, the set $\{\calT(p\alpha)\mid p\in Q, \alpha\in\Gamma^*\}$
is closed {w.r.t.}~reachability in $\calL^{\ltsact}_{\calG_M}$
(if $\calT(p\alpha)\gt{a}F$ in  $\calL^{\ltsact}_{\calG_M}$, then 
$F=\calT(q\beta)$ where  $p\alpha\gt{a}q\beta$ in $\calL_M$).

In fact, we have not allowed $\varepsilon$-rules
$A(x_1,\dots,x_m)\gt{\varepsilon}E$ in our definition of first-order
grammars. We would consider a variant of so called \emph{weak
bisimilarity} in such a case, which is undecidable in general
(see, e.g.,~\cite{DBLP:journals/jacm/JancarS08} for a further discussion).

At the end of Section~\ref{sec:prelim} we mention
\emph{restricted PDAs} where \emph{$\varepsilon$-rules}
$pY\gt{\varepsilon}q\alpha$ can be only \emph{popping}, i.e.
$\alpha=\varepsilon$ in such rules, and \emph{deterministic} (or 
\emph{having no alternative}), which
means that if there is a rule $pY\gt{\varepsilon}q$ in $\Delta$
then there is no other rule with the left-hand side $pY$ (of the form
$pY\gt{a}q'\alpha$ where $a\in\Sigma\cup\{\varepsilon\}$).
We define the \emph{stable configurations} as $p\varepsilon$ and
$pY\alpha$ where there is no rule $pY\gt{\varepsilon}..$ in $\Delta$;
for the restricted PDAs we have that any unstable configuration
$p\alpha$ only allows to perform a finite sequence of
$\varepsilon$-transitions that reaches a stable configuration. Hence
it is natural to restrict the attention to the ``visible'' transitions 
$p\alpha\gt{a}q\beta$ ($a\in\Sigma$) between stable configurations;
such transitions might encompass sequences of $\varepsilon$-steps.
In defining the grammar $\calG_M$ we can naturally avoid the explicit use of deterministic
popping $\varepsilon$-transitions, by ``preprocessing'' them: in our
inductive definition of
$\calT(p\alpha)$ (and $\calT(p\alpha x)$) we add the following
item: if $pY$ is unstable, since there is a rule $pY\gt{\varepsilon}q$,
then $\calT(pY\alpha)=\calT(q\alpha)$.
\begin{figure}[ht]
\centering
\includegraphics[scale=0.43]{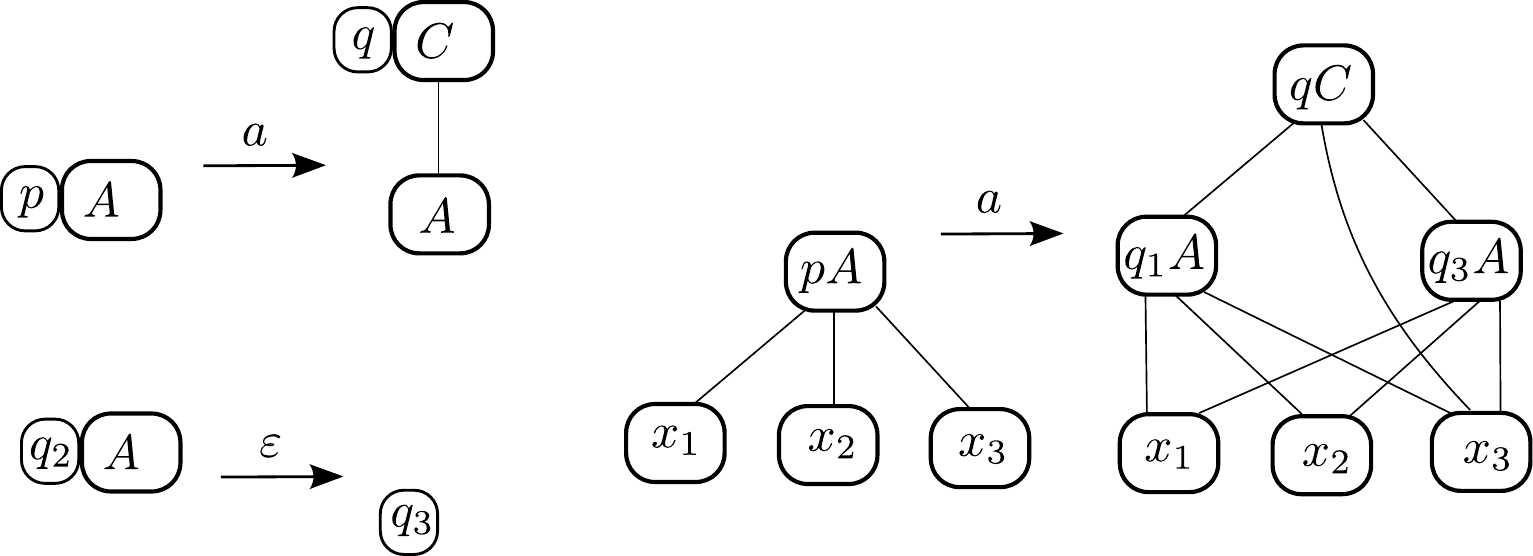}
	\caption{Deterministic popping $\varepsilon$-transitions are
	``preprocessed''}\label{fig:swalloweps}
\end{figure}
Fig.~\ref{fig:swalloweps} (right) shows the grammar-rule
$\calT(pAx)\gt{a}\calT(qCAx)$ (arising from the PDA-rule
$pA\gt{a}qCA$), when $Q=\{q_1,q_2,q_3\}$ and 
there is a PDA-rule $q_2A\gt{\varepsilon}q_3$, while $q_1A$, $q_3A$
are stable.
Such preprocessing causes that the term $\calT(p\alpha)$
can have branches of varying lengths.

\subsection*{Normalization of grammars}

We call a \emph{grammar} $\calG=(\calN,\act,\calR)$ \emph{normalized}
if for each $A\in\calN$ and each $i\in[1,\arity(A)]$
there is
a (``sink'') word $w_{(A,i)}\in\calR^+$
such that $A(x_1,\dots,x_{\arity(A)})\trans{w_{(A,i)}}x_i$.

For any grammar $\calG=(\calN,\act,\calR)$ we can find some
words $w_{(A,i)}$ or find out their non-existence, for all 
$A\in\calN$ and $i\in[1,\arity(A)]$, as shown below. For technical 
convenience we will also find some words  $w_{(E',x_i)}\in\calR^*$ satisfying
$E'\trans{w_{(E',x_i)}}x_i$ for subterms $E'$ of the rhs (right-hand
sides)
$E$ of the rules $A(x_1,\dots,x_m)\gt{a}E$ in $\calR$.

We put $w_{(x_i,x_i)}=\varepsilon$ (for subterms $x_i$ of the rhs), while
all other $w_{(E',x_i)}$ and all $w_{(A,i)}$ are undefined in the beginning.
Then we repeatedly define \emph{so far undefined} 
$w_{(A,i)}$ or $w_{(E',x_i)}$ by applying 
the following constructions, as long 
as possible:
\begin{itemize}
\item
 put $w_{(A,i)}=r\,w_{(E,x_i)}$ if
	there is a rule 
		$r:A(x_1,\dots,x_m)\gt{a}E$ and $w_{(E,x_i)}$ is
		defined;
	\item 	put 	$w_{(E',x_i)}=w_{(A,j)}\,w_{(E'',x_i)}$
		if $\termroot(E')=A$, $E''$ is the $j$th root-successor
		in $E'$, and $w_{(A,j)}$, $w_{(E'',x_i)}$ are defined.
\end{itemize}		
The correctness is obvious. The process could be modified to find
some \emph{shortest} $w_{(A,i)}$ (and $w_{(E',x_i)}$) that exist but this is
not important here.
For any pair $(A,i)$ for which $w_{(A,i)}$ has remained undefined
such word obviously does not exist, hence the $i$th root-successor $G_i$ 
of any term
$A(G_1,\dots,G_m)$ is
``non-exposable'' and thus 
plays ``no role'' (not affecting the bisim-class of
$A(G_1,\dots,G_m)$). We will now show a safe removal of such non-exposable
root-successors.

\begin{figure}[ht]
\centering
\includegraphics[scale=0.45]{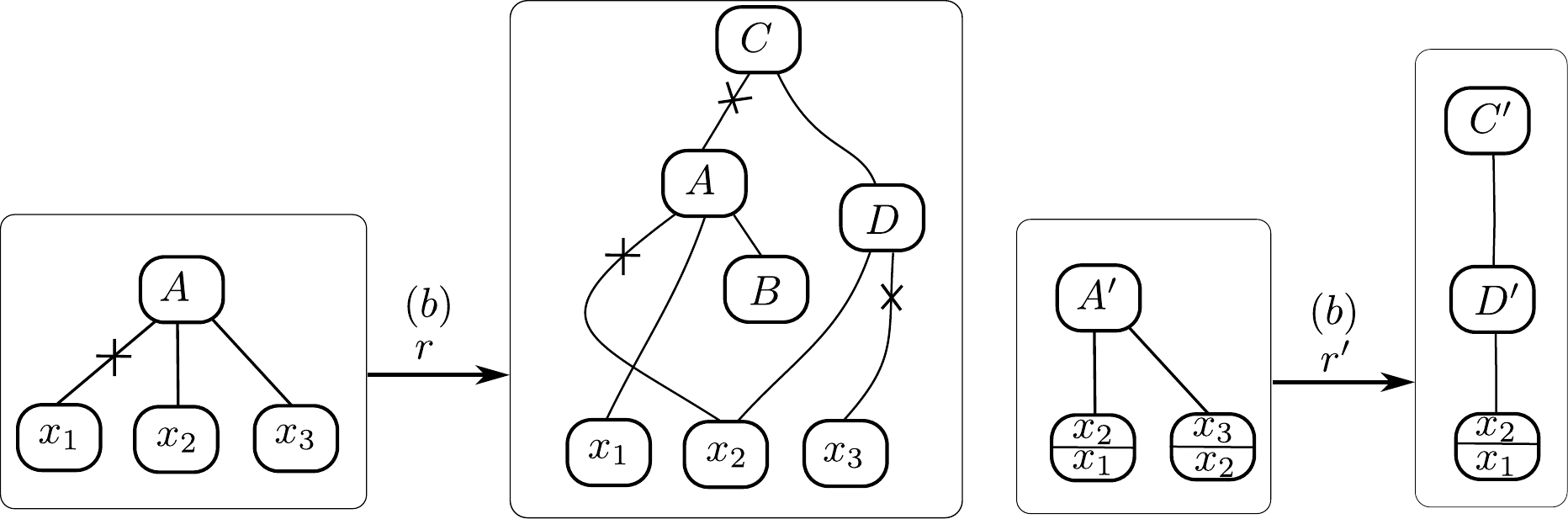}
	\caption{Modifying (cutting) a rule $r$, when $w_{(A,1)}$, $w_{(C,1)}$, 
	and $w_{(D,2)}$ do not exist}\label{fig:cutnormaliz}
\end{figure}

For $\calG=(\calN,\act,\calR)$
we put $\calG'=(\calN',\act,\calR')$ where the sets
$\calN'=\{A'\mid A\in\calN\}$ and $\calR'=\{r'\mid r\in\calR\}$
 are defined below. 
For $A\in\calN$, we put 
\begin{center}
$\sink(A)=\{i\in[1,\arity(A)]\mid$ there is
some $w_{(A,i)} \}$, and 
	$\arity(A')=|\sink(A)|$.
\end{center}
 We define the mapping
 $\cut:\trees_\calN\rightarrow\trees_{\calN'}$ by the following
 structural induction:
\begin{enumerate}
	\item		
$\cut(x_i)=x_i$\,;
\item
$\cut(A(G_1,G_2,\dots,G_{m}))=
		A'\,(\cut(G_{i_1}),\cut(G_{i_2}),\dots,\cut(G_{i_{m'}}))$,
\\
where
$1\leq i_1<i_2<\cdots <i_{m'}\leq m$ and 
		$\{i_1,i_2,\dots,i_{m'}\}=\sink(A)$.
\end{enumerate}
The set of rules $\calR'=\{r'\mid r\in\calR\}$ is defined as follows:
\begin{center}
for $r:A(x_1,\dots,x_m)\gt{a}E$ we put
 $r':\cut(A(x_1,\dots,x_m))\sigma\gt{a}\cut(E)\sigma$
\end{center}
	where 
 $\sigma=\{(x_{i_1},x_1), (x_{i_2},x_2),\dots,
(x_{i_{m'}},x_{m'})\}$ for 	$\{i_1,i_2,\dots,i_{m'}\}=\sink(A)$.
(Fig.~\ref{fig:cutnormaliz} depicts the transformation of 
$r:A(x_1,x_2,x_3)\gt{b}C(A(x_2,x_1,B),D(x_2,x_3))$ to
$r':A'(x_2,x_3)\sigma\gt{b}C'(D'(x_2))\sigma$ where  
$\sigma=\{(x_2,x_1), (x_3,x_2)\}$.)
We note that for each variable $x_i$ occurring in $\cut(E)$
we must have $i\in\sink(A)=\{i_1,i_2,\dots,i_{m'}\}$; hence $\sigma$
yields a one-to-one renaming of ``place-holders'' 
in 
$\cut(A(x_1,\dots,x_m))\gt{a}\cut(E)$.

It is easy to check that $\cut$ maps $\trees_\calN$ onto
$\trees_{\calN'}$, and that
 $G\gt{r}H$ in $\calL^{\ltsrul}_\calG$ 
implies 
$\cut(G)\gt{r'}\cut(H)$ in $\calL^{\ltsrul}_{\calG'}$;
moreover, if $G'\gt{r'}H'$ in  $\calL^{\ltsrul}_{\calG'}$ and
$G\in\cut^{-1}(G')$, then there is $H\in\cut^{-1}(H')$ such that
$G\gt{r}H$ in $\calL^{\ltsrul}_{\calG}$.

Grammar $\calG'$ is normalized: 
if $\sink(A)=\{i_1,i_2,\dots,i_{m'}\}$ then for each $j\in[1,m']$ we have 
$A(x_1,\dots,x_{m})\trans{w_{(A,i_j)}}x_{i_j}$, and thus
$\cut(A(x_1,\dots,x_{m}))\trans{(w_{(A,i_j)})'}\cut(x_{i_j})$,
{i.e.}~$A'(x_{i_1},\dots,x_{i_{m'}})\trans{(w_{(A,i_j)})'}x_{i_j}$,
where $w'$ arises
from $w$ by replacing each element $r$ with $r'$.

We also have that
the set $\{(F,\cut(F))\mid F\in\trees_\calN\}$ 
is a bisimulation 
in the union of $\calL^{\ltsact}_\calG$ and
$\calL^{\ltsact}_{\calG'}$; hence $F\sim\cut(F)$,
and $E_0$ is bisim-finite in 
 $\calL^{\ltsact}_{\calG}$ iff $\cut(E_0)$ is bisim-finite in 
 $\calL^{\ltsact}_{\calG'}$.
(We can also note that the quotient-LTS
$(\calL^{\ltsact}_{\calG})_{\equiv_{\cut}}$ 
is isomorphic with  $\calL^{\ltsact}_{\calG'}$, and 
that the bisimilarity quotients of 
$\calL^{\ltsact}_{\calG}$ and $\calL^{\ltsact}_{\calG'}$ are the same,
up to isomorphism.)

\subsection*{Unification of nonterminal arities}

In the proof of Theorem~\ref{th:decidsemfinit}
we used $m$ for denoting the arity of each
nonterminal in the considered normalized grammar,
instead of using $m_A$ for $\arity(A)$. 
This was not crucial, since it would be straightforward 
to modify the relevant arguments in the proof,
but we also mentioned that we could ``harmlessly'' achieve the uniformity of
nonterminal arities by a construction,  while keeping the
adjusted grammar normalized. We now sketch such a construction.

\begin{figure}[ht]
\centering
\includegraphics[scale=0.35]{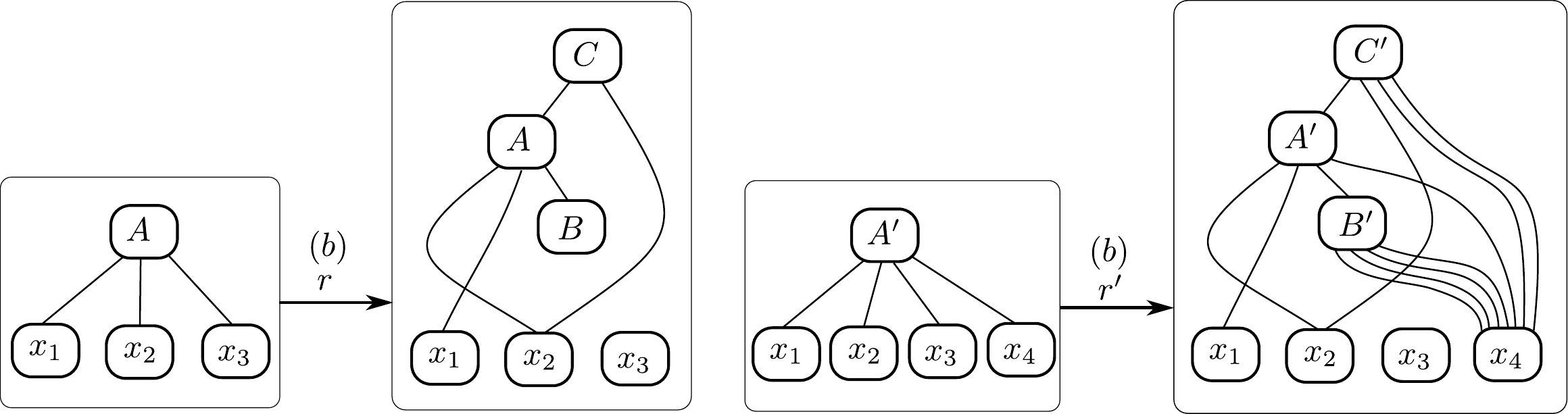}
	\caption{Padding a rule in $\calR$ (left) 
	with $x_4$, when $m=3$}\label{fig:padrule}
\end{figure}

\begin{figure}[ht]
\centering
\includegraphics[scale=0.4]{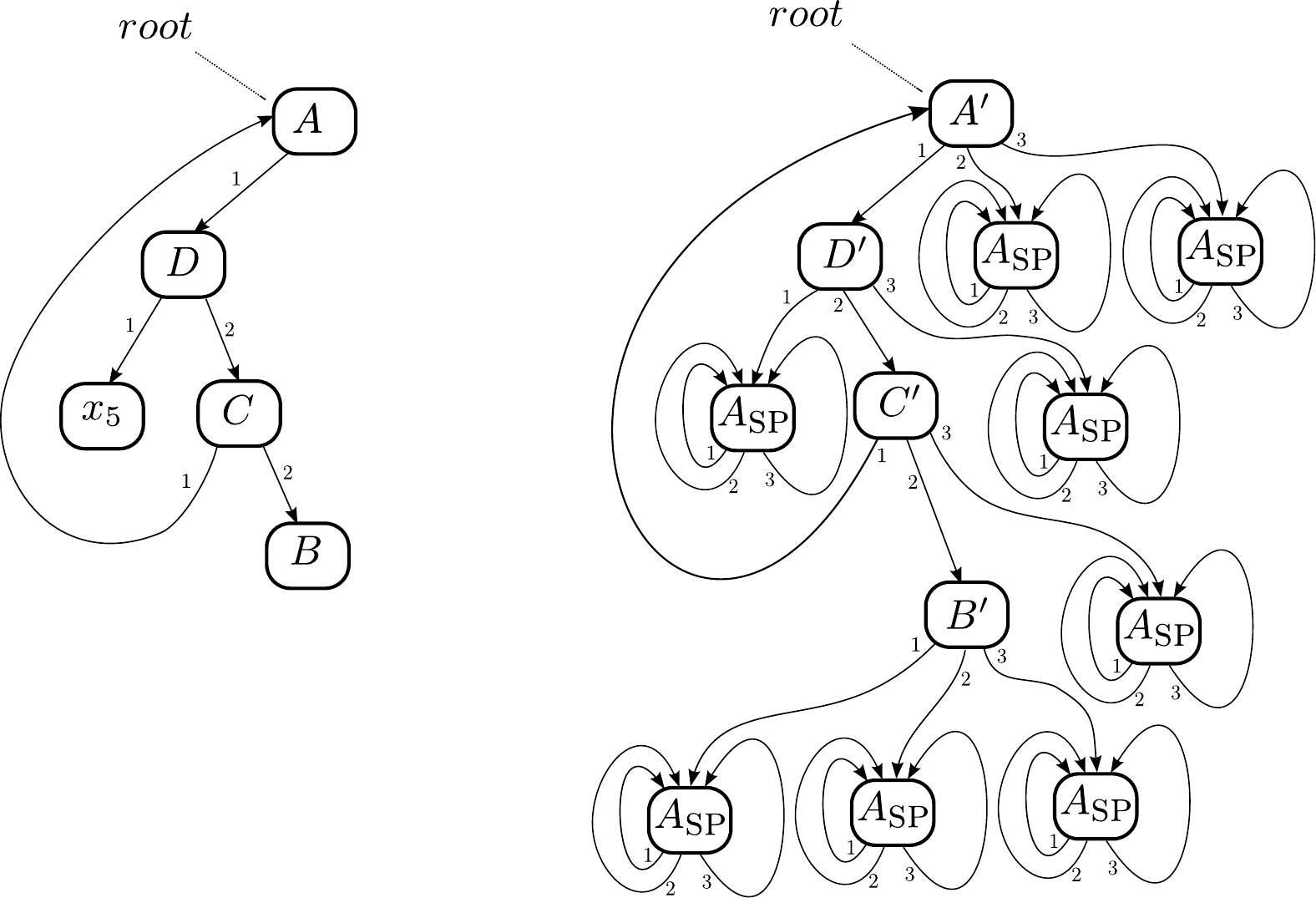}
	\caption{A term $F\in\trees_\calN$ (left) and $\pad_\spec(F)$
	(right), when $m=2$}\label{fig:padterms}
\end{figure}

Suppose $\calG=(\calN,\act,\calR)$ is normalized and
 the arities of
nonterminals are not all the same; let $m$ be the maximum arity.
If there are no nullary nonterminals, then the arities can be unified
to $m$ by a straightforward ``padding with superfluous copies
of root-successors''. But we will pad with a special (infinite
regular) term, which 
 handles the case
of nullary nonterminals as well. (This is illustrated in
Figures~\ref{fig:padrule} and~\ref{fig:padterms}.)

We define the grammar $\calG'=(\calN',\act',\calR')$ 
where 
$\calN'=\{A'\mid A\in\calN\}\cup\{\Aspec\}$,
$\act'=\act\cup\{\aspec\}$, and 
$\calR'=\{r'\mid r\in\calR\}\cup\Rspec$ as 
defined below. Each nonterminal in $\calN'$, including the special
nonterminal $\Aspec$, has arity $m{+}1$.
The set $\Rspec$ (of the rules with the special added action
$\aspec$) contains the following rules:
\begin{itemize}
	\item
	$\Aspec(x_1,\dots,x_{m+1})\trans{\aspec}x_i$, for all 
		$i\in[1,m{+}1]$;
	\item
	$A'(x_1,\dots,x_{m+1})\trans{\aspec}x_i$, for all
 $A\in\calN$ and $i\in[\arity(A){+}1,m{+}1]$.
\end{itemize}	
The definition of $\calR'=\Rspec\cup\{r'\mid r\in\calR\}$
is finished by the following point (see Fig.~\ref{fig:padrule}):
\begin{itemize}
	\item
for $r:A(x_1,\dots,x_{\arity(A)})\gt{a}E$ we put
		$r':A'(x_1,\dots,x_{m+1})\gt{a}\pad(E,x_{m+1})$. 
\end{itemize}
The expression $\pad(E,x_{m+1})$
 is clarified by the following 
inductive definition of  $\pad(F,H)$
(padding $F\in\trees_\calN$ with certain $H$):
\begin{enumerate}
	\item		
$\pad(x_i,H)=x_i$\,;
\item
$\pad(A(G_1,\dots,G_{m'}),H)=
 A'\,(\pad(G_1,H),\dots,\pad(G_{m'},H), H,\dots,H)$,
\\
where $m'=\arity(A)$, and $m{+}1{-}m'$ copies of $H$ are used to
		``fill''
		the arity $m{+}1$ of $A'$.
\end{enumerate}
Besides the above case $\pad(E,x_{m+1})$ 
we use the definition of $\pad(F,H)$ also for
$H=\Espec$, i.e., for the special (infinite regular) term 
\begin{center}
$\Espec=\Aspec(x_1,\dots,x_{m+1})\sigma^\omega$ where 
$x_i\sigma=\Aspec(x_1,\dots,x_{m+1})$ for all $i\in[1,m{+}1]$. 
\end{center}
(In Fig.~\ref{fig:padterms} 
there are several copies of the least presentation of  $\Espec$ when
$m=2$.)
The behaviour of $\Espec$ is trivial: its only outgoing transition in
$\calL^{\ltsact}_{\calG'}$ is
the loop $\Espec\gt{\aspec}\Espec$.

We define the mapping
 $\pad_\spec:\trees_\calN\rightarrow\trees_{\calN'}$ by
$\pad_\spec(F)=\pad(F,\Espec)\sigma_\spec$
where $x_i\sigma_\spec=\Espec$ for each variable $x_i$.
(The support of $\sigma_\spec$
is infinite but this causes no
problem.)
Hence there are no variables in  $\pad_\spec(F)$.
(Fig.~\ref{fig:padterms} shows an example. We note that if the nullary
nonterminal $B$ happens
to be dead in $\calL^{\ltsact}_{\calG}$, then $B'\sim \Espec$  in 
$\calL^{\ltsact}_{\calG'}$; this is the reason for replacing the
variables with $\Espec$.)

The mapping $\pad_\spec$ is injective (but not onto $\trees_{\calN'}$)
and the following conditions obviously hold: 
\begin{itemize}
\item		
if $G\gt{r}H$ (in $\calL^{\ltsrul}_\calG$) then 
$\pad_\spec(G)\gt{r'}\pad_\spec(H)$ (in $\calL^{\ltsrul}_{\calG'}$);
\item
if $\pad_\spec(G)\gt{r'}H'$ then there is $H$ such that
$\pad_\spec(H)=H'$ and $G\gt{r}H$. 
\end{itemize}
The rules in $\calR_\spec$ guarantee 
that $\calG'$ is normalized
(if $\calG$ is normalized), and they also
induce that  
the special action $\aspec$ is enabled in any term
$\pad_\spec(G)$ (in $\calL^{\ltsact}_{\calG'}$); moreover, 
 $\pad_\spec(G)\trans{\aspec}H'$ entails that $H'=\Espec$.

We now note that any set $\calB\subseteq \trees_\calN\times\trees_\calN$
is a bisimulation in $\calL^{\ltsact}_{\calG}$ iff
$\calB'=\{(\Espec,\Espec)\}\cup\{(\pad_\spec(F),\pad_\spec(G))\mid
(F,G)\in\calB)\}$ is a bisimulation in $\calL^{\ltsact}_{\calG'}$.
We deduce that  
$E\sim F$ in $\calL^{\ltsact}_{\calG}$ iff  $\pad_\spec(E)\sim
\pad_\spec(F)$ in $\calL^{\ltsact}_{\calG'}$, and $E_0$ is bisim-finite in 
$\calL^{\ltsact}_{\calG}$ iff 
$\pad_\spec(E_0)$ is bisim-finite in 
$\calL^{\ltsact}_{\calG'}$.

\subparagraph*{Author's acknowledgements.}
This research was mostly carried out when I was affiliated
with Techn. Univ. Ostrava and 
supported by the Grant Agency of the Czech Rep.,
project GA\v{C}R:15-13784S.
I also thank Stefan G\"oller for drawing my attention to
the decidability question for regularity of pushdown processes,
for discussions about some
related works (like~\cite{DBLP:journals/jacm/Valiant75}), and for
detailed comments on a previous version of this paper.
My thanks also go to anonymous reviewers for their helpful comments that 
have also prompted me to further simplifications of the proof.

\bibliographystyle{elsarticle-num}
\bibliography{pj-mfcs-16}

\begin{thebibliography}{10}
\expandafter\ifx\csname url\endcsname\relax
  \def\url#1{\texttt{#1}}\fi
\expandafter\ifx\csname urlprefix\endcsname\relax\def\urlprefix{URL }\fi
\expandafter\ifx\csname href\endcsname\relax
  \def\href#1#2{#2} \def\path#1{#1}\fi

\bibitem{DBLP:journals/iandc/Stearns67}
R.~E. Stearns, A regularity test for pushdown machines, Information and Control
  11~(3) (1967) 323--340.

\bibitem{DBLP:journals/jacm/Valiant75}
L.~G. Valiant, Regularity and related problems for deterministic pushdown
  automata, J. ACM 22~(1) (1975) 1--10.

\bibitem{Senizergues:TCS2001}
G.~S\'{e}nizergues, {L(A)=L(B)?} {D}ecidability results from complete formal
  systems, Theoretical Computer Science 251~(1--2) (2001) 1--166.

\bibitem{Milner1989}
R.~Milner, Communication and Concurrency, Prentice-Hall, Inc., Upper Saddle
  River, NJ, USA, 1989.

\bibitem{Srba:Roadmap:04}
J.~Srba, Roadmap of infinite results, in: Current Trends In Theoretical
  Computer Science, The Challenge of the New Century, Vol.~2, World Scientific
  Publishing Co., 2004, pp. 337--350, updated version at
  http://users-cs.au.dk/srba/roadmap/.

\bibitem{Seni05}
G.~S\'enizergues, The bisimulation problem for equational graphs of finite
  out-degree, SIAM J.Comput. 34~(5) (2005) 1025--1106.

\bibitem{BGKM12}
M.~Benedikt, S.~G{\"o}ller, S.~Kiefer, A.~S. Murawski, Bisimilarity of pushdown
  automata is nonelementary, in: Proc. LICS 2013, IEEE Computer Society, 2013,
  pp. 488--498.

\bibitem{DBLP:journals/toct/Schmitz16}
S.~Schmitz, Complexity hierarchies beyond elementary, {TOCT} 8~(1) (2016) 3.

\bibitem{DBLP:conf/fossacs/Jancar14}
P.~Jan\v{c}ar, Equivalences of pushdown systems are hard, in: Proc. FOSSACS
  2014, Vol. 8412 of LNCS, Springer, 2014, pp. 1--28.

\bibitem{Stir:DPDA:prim}
C.~Stirling, Deciding {DPDA} equivalence is primitive recursive, in: Proc.
  ICALP'02, Vol. 2380 of LNCS, Springer, 2002, pp. 821--832.

\bibitem{Kucera10}
A.~Ku\v{c}era, R.~Mayr, On the complexity of checking semantic equivalences
  between pushdown processes and finite-state processes, Inf. Comput. 208~(7)
  (2010) 772--796.

\bibitem{DBLP:conf/fsttcs/BroadbentG12}
C.~H. Broadbent, S.~G{\"o}ller, On bisimilarity of higher-order pushdown
  automata: Undecidability at order two, in: FSTTCS 2012, Vol.~18 of LIPIcs,
  Schloss Dagstuhl - Leibniz-Zentrum f{\"u}r Informatik, 2012, pp. 160--172.

\bibitem{DBLP:conf/icalp/Jancar14}
P.~Jan\v{c}ar, Bisimulation equivalence of first-order grammars, in: Proc.
  ICALP'14 (II), Vol. 8573 of LNCS, Springer, 2014, pp. 232--243.

\bibitem{DBLP:journals/tcs/Courcelle95}
B.~Courcelle, The monadic second-order logic of graphs {IX:} machines and their
  behaviours, Theor. Comput. Sci. 151~(1) (1995) 125--162.

\bibitem{DBLP:conf/fossacs/KnapikNU02}
T.~Knapik, D.~Niwinski, P.~Urzyczyn, Higher-order pushdown trees are easy, in:
  Proc. FOSSACS 2002, Vol. 2303 of LNCS, Springer, 2002, pp. 205--222.

\bibitem{DBLP:conf/lics/Ong15}
L.~Ong, Higher-order model checking: An overview, in: Proc. LICS 2015, {IEEE}
  Computer Society, 2015, pp. 1--15.

\bibitem{DBLP:journals/siglog/Walukiewicz16}
I.~Walukiewicz, Automata theory and higher-order model-checking, {ACM SIGLOG}
  News 3~(4) (2016) 13--31.

\bibitem{CourcelleHandbook}
B.~Courcelle, Recursive applicative program schemes, in: J.~van Leeuwen (Ed.),
  Handbook of Theoretical Computer Science, vol. B, Elsevier, MIT Press, 1990,
  pp. 459--492.

\bibitem{DBLP:journals/jacm/JancarS08}
P.~Jan\v{c}ar, J.~Srba, Undecidability of bisimilarity by defender's forcing,
  J. ACM 55~(1).

\bibitem{DBLP:journals/corr/abs-1303-0780}
P.~Jan\v{c}ar, J.~Srba, Note on undecidability of bisimilarity for second-order
  pushdown processes, CoRR abs/1303.0780.

\end{thebibliography}

\end{document}